\documentclass[12pt,prd,aps,amssymb,amsmath,tightenlines,showpacs]{article}
\usepackage[utf8]{inputenc}
\usepackage[a4paper,
top=1cm,bottom=3cm,left=2.5cm,right=2.5cm]{geometry}
\usepackage{amssymb}
\usepackage{amsmath}
\usepackage{amsthm}
\usepackage{tipa}
\usepackage{latexsym} 
\usepackage{graphicx}
\usepackage{slashed}
\usepackage{bbold}
\usepackage{cancel}
\usepackage{comment}
\usepackage{color}
\usepackage{graphicx,epsfig,color}
\usepackage{comment}
\usepackage{cite}
\usepackage{tikz}
\usepackage[compat=1.1.0]{tikz-feynman}

\newcommand{\be}{\begin{equation}}
\newcommand{\ee}{\end{equation}}
\newcommand{\bea}{\begin{eqnarray}}
\newcommand{\eea}{\end{eqnarray}}
\newcommand{\vv}{``}

\def\nn{\nonumber}

\begin{document}
	\graphicspath{{FIGURE/}}
	\topmargin=-2cm

\begin{center} {\Large{\bf Dimensional regularization, Wilsonian RG,  \\ and the Naturalness/Hierarchy problem}}

\vspace*{0.8 cm}
	{ Carlo Branchina$^{*,\,a}$ \let\thefootnote\relax\footnote{$^{*}$cbranchina@lpthe.jussieu.fr},
Vincenzo Branchina$^{\dagger,\,b}$ \let\thefootnote\relax\footnote{$^\dagger$branchina@ct.infn.it},
Filippo Contino$^{\ddagger,\,b}$ \footnote{$^\ddagger$filippo.contino@ct.infn.it} and Neda Darvishi$^{\star,\,c,d}$ \footnote{$^\star$neda.darvishi@itp.ac.cn}}
		
				\vspace*{0.4cm}
		
		{${}^{a}$\it Laboratoire de Physique Th\'eorique et Hautes Energies (LPTHE), UMR 7589,\\
Sorbonne Universit\'e et CNRS, 4 place Jussieu, 75252 Paris Cedex 05, France}
		
				\vspace*{0.4cm}
		
		{${}^b$\it Department of Physics, University of Catania, 
			and INFN,} \\
		\vspace*{0.02cm}
		{\it Via Santa Sofia 64, I-95123 
			Catania, Italy}\\
	
		\vspace*{0.4cm}
	
	{${}^c$\it	Department of Physics and Astronomy, Michigan State University,\\ East Lansing, MI
	48824,USA}

	\vspace*{0.4cm}

{${}^d$\it Institute of Theoretical Physics, Chinese Academy of Sciences, Beijing 100190, China}
		
		\vspace*{1 cm}

		{\LARGE Abstract}\\
		
\end{center}

While it is usually stated that dimensional regularization (DR) has no direct physical interpretation, consensus has recently grown on the idea that it might be endowed with special physical properties that would provide the mechanism that solves the naturalness/hierarchy problem. Comparing direct Wilsonian calculations with the corresponding DR ones, we find that DR indeed has a well-defined physical meaning, and we point out its limitations. In particular, our results show that DR cannot provide the solution to the naturalness/hierarchy problem. The absence of too large corrections to the Higgs boson mass is due to a secretly realized fine-tuning, rather than special physical properties of DR. We also investigate these issues within the Wilsonian RG framework and, by comparison with the usual perturbative RG analysis,  we show that several popular proposals for the resolution of the  problem, commonly considered as physical mechanisms free of fine-tuning, again secretly implement the tuning. 

\setcounter{footnote}{0}

\section{Introduction}
\label{intro}

The Standard Model (SM) of particle physics is a very successful theory, and the discovery of the Higgs boson\cite{ATLAS:2012yve, CMS:2012qbp} is one the most important findings of the last years. However it is not a  complete theory, and the search for physics beyond the Standard Model (BSM) is one of the strongest driving force of present experimental and theoretical physics. Several fundamental unsolved questions (dark matter, matter-antimatter asymmetry, neutrino masses, the flavour and the strong CP-problem, the problem with the unification of the gauge couplings) urge us to find the way beyond the SM. Among them, the naturalness and hierarchy problems.

Any QFT that contains scalar fields is  confronted with the naturalness problem. It is formulated in different (equivalent) ways, the essential point being that the quantum corrections to the mass of a scalar field are typically proportional to the \vv highest mass scale'' of the theory. When this scale is too large, we have to resort to an \vv unnatural" tuning of the mass parameter, a \vv fine-tuning". 

One way to formulate the problem is
as follows (see for instance\cite{Giudice:2013yca}).
If the higher energy model assumed to embed the SM contains a field of large mass $M$ coupled to the Higgs field $H(x)$, its mass $m_H$ 
receives corrections  proportional to $M$ ($ \gg m_H$).  
As an example, we can consider GUT models, 
that contain scalar fields $\phi$ with masses $M \gg m_H$. These fields are coupled to $H$ through terms of the kind  $\lambda_{\phi} H^2\phi^2$, so that $m_H$ receives corrections as ($\mu$ is the subtraction or 't Hooft scale)
\be\label{GUT}
\Delta m_H^2 \sim \lambda_{\phi} \,M^2\, {\rm ln}\frac {M^2}{\mu^2}\,.
\ee 
Similarly, in a supersymmetric extension of the SM,
where SUSY is broken by 
a large stop mass $\widetilde m_t \gg m_H$, the Higgs mass receives a large correction ($y_t$ is the top Yukawa coupling)
\be\label{stop}
\Delta m_H^2 \sim y_{t}\, \widetilde m_t^2\, {\rm ln}\frac {\widetilde m_t^2}{\mu^2}\,.
\ee 

The same problem can be formulated in a Wilsonian Effective Field Theory (WEFT) framework\cite{Susskind:1978ms}, where a QFT is defined with a built-in cut-off $\Lambda$, the scale above which the theory loses its validity and has to be replaced with a (not better specified) higher energy theory. In this framework, the \vv highest mass scale'' of the SM is $\Lambda$, and $\Delta m_H^2$ takes the form 
\be\label{cut-off}
\Delta m_H^2 \sim {\rm \alpha}\, \Lambda^2\,,
\ee 
where $\alpha$ is a combination of coupling constants. 

Despite their apparent difference,  Eqs.\,(\ref{GUT})-(\ref{stop}) on the one side, and (\ref{cut-off}) on the other one, have the same physical content. In fact, the SM 
is an effective \vv low-energy" theory valid up to a certain ``new physics" scale\,\footnote{Barring the possibility that the SM is the Theory of Everything.}, irrespectively of being such a scale an ultraviolet (UV) cut-off $\Lambda$ (that in particular could be the Planck mass  $M_P$), or the mass of a
heavy GUT scalar or of a heavy stop.

From the low-energy perspective, i.e.\,\,from the perspective of the SM alone, we do not know its UV completion.
The Higgs mass $m_H^2$ receives a \vv quadratic correction'' (in the form (\ref{GUT}) or (\ref{stop}) or (\ref{cut-off})), as a generic \vv left-over'' of the higher energy theory that embeds the SM. The reason is that in the low-energy theory (the SM in this case) there is no symmetry  that protects $m_H^2$ from getting such a large 
contribution. 

Referring to the above example of a SUSY embedding, it is often stated that, even though (due to supersymmetry) there are no \vv quadratic divergences'', still there is a large correction to $m^2_H$ (see (\ref{stop})), concluding that the naturalness problem is not related to the occurrence of quadratic divergences, but rather to the presence of high mass scales in the theory\cite{Giudice:2013yca,deGouvea:2014xba}.
However, from the viewpoint of the lower energy theory (the SM), ${m^2_H}$\,  
{\it does get}\,\, a \vv quadratically divergent'' correction, that is $\widetilde m_t^2$, the scale above which the SM is no longer valid. This is the  physical meaning of  \vv quadratically divergent" correction. Below $\widetilde m_t$ the theory is the SM, not its supersymmetric extension. Therefore we should not refer to the absence of quadratic divergences in the SUSY theory, but rather to the presence of quadratic divergences in the SM. 

If we generically indicate  with $\Lambda$ the scale above which the SM has to be replaced by a higher energy theory (the \vv highest mass scale" of the SM), the radiative corrections 
to $m^2_H$ are proportional to  $\Lambda^2$. 
We stress that  $\Lambda$ is not a cut-off to be sent to infinity, but rather a physical scale above which the physics cannot be described in terms of the low-energy SM degrees of freedom (dof). 
The \vv quadratic correction'' to $m^2_H$ is considered as \vv unnatural'' in the sense that it is too large: $\Delta m_H^2 \sim \Lambda^2$.
Therefore, the relevant question is  whether there exists a \vv physical mechanism" (a symmetry, a dynamical mechanism, ...) that could free $m_H^2$ from such a large correction, allowing to get
\be\label{natur}
m^2_H \ll \Lambda^2\,.
\ee
In the absence of such a mechanism, we are lead to resort to an ad hoc \vv fine-tuning'' of the mass.

A great progress in our understanding of (renormalisation in) QFTs 
comes from the Wilson's lesson, that relies on the deep connection between QFTs and statistical physics\,\footnote{From the theory of critical phenomena we know that the critical regime is reached when the correlation 
length\,$\xi$\,among statistical fluctuations becomes much larger than the inter-atomic distance $a$. For a ferromagnet this happens when the temperature $T$ approaches the critical temperature  $T_c$, and for $T$ close to $T_c$ we have $\xi \sim |T-T_c|^{-\nu}$, where $\nu (>0)$ is the appropriate critical exponent. QFTs and critical phenomena are connected through the correspondence $m_H \to \frac{1}{\xi}$ \,,\, $\Lambda \to \frac{1}{a}$, and the requirement (\ref{natur}) in QFT corresponds to the tuning of the statistical system toward the critical regime.}.
In the Wilsonian framework, first the fluctuation modes are eliminated within a tiny shell, then a tuning of the parameters towards the critical region is realized. Iterating this procedure, the renormalized theory is finally obtained. For an interesting implementation of these ideas in the Hamiltonian framework see\cite{Glazek:1994qc}.

This puts the renormalization of a QFT on a totally different perspective than a mere affair of \vv cancellation of divergences". However, it gives no clue on the {\it physical mechanism} that drives the system toward the \vv critical regime''  $m^2_H \ll \Lambda^2$. In this respect, there is a profound difference between critical phenomena and QFTs. While the mechanism that drives the statistical system towards the critical regime is well known (in the case of a ferromagnet, this is the tuning of the temperature towards $T_c$), for QFTs we do not know what triggers the system towards the critical regime, i.e. towards the renormalized theory. 

The above considerations indicate what the \vv physical" way of posing the naturalness problem should be. The mass $m_H^2$ at the scale $\Lambda$ is \vv naturally'' expected to be  $m_H^2(\Lambda) \sim \Lambda^2$. Which physical mechanism drives the SM toward the critical regime  $m^2_H (v) \ll \Lambda^2\,$ ($v$ is the Fermi scale)? 

In other words, which physical mechanism introduces such an unnatural hierarchy among physical scales?

Traditional approaches, as supersymmetry and/or composite models, have to cope with the unfriendly constraints that come from the LHC results: the compositeness scale or the SUSY breaking scale should be just around the corner, in the TeV regime, but no sign of new physics has been observed so far. This leads several authors to speculate that the SM could be valid up to some very high energy scale, the Planck scale $M_P$ or so. If this is the case, alternative approaches to the naturalness problem have to be envisaged.

In this respect, some authors  consider (classical) conformal extensions of the SM, where the quantum fluctuations break the conformal symmetry only softly. 
By taking models with no intrinsic mass scale, i.e.\,containing only operators of dimension four, and calculating the quantum fluctuations using  dimensional regularization (DR), only a logarithmic breaking of the conformal symmetry is realized, and small masses appear through a Coleman-Weinberg mechanism\,\footnote{Conformal extensions of the SM have also been advocated elsewhere\cite{tHooft:2011aa,tHooft:2016uxd}, in a somehow different perspective: the couplings should not run with the scale (vanishing $\beta$ functions), and the model should have  enough constraints so that all the parameters should be fixed (predicted). This very ambitious program, however, has not yet found realistic implementations.}\cite{Meissner:2006zh,Salvio:2014soa,Boyle:2011fq,Alexander-Nunneley:2010tyr,Carone:2013wla,Farzinnia:2013pga,Ghilencea:2015mza,Ghilencea:2016dsl,Guo:2014bha,Kawamura:2013kua,Foot:2007iy,Meissner:2007xv,Oda:2018zth,Ghilencea:2016ckm,Heikinheimo:2013fta,Mooij:2018hew,Shaposhnikov:2008xi,Bezrukov:2014ipa,Bezrukov:2007ep,Bars:2013yba,Steele:2013fka,Wang:2015cda}.

The central assumption of these approaches is that DR plays a special role in defining the renormalized theory (see for instance \cite{Meissner:2006zh}), grasping an element of truth that is missed by the Wilson's theory.
The two methods are regarded as physically different (see for instance\cite{Salvio:2014soa}): the Wilsonian one needs a \vv fine-tuning'', while DR seems to be dispensed of it. Even more,  the Wilsonian paradigm is 
downgraded to a ``computational technique'' that {\it improperly}
insists in giving a physical meaning to the elimination of momentum shells in the construction of the effective action, and the naturalness problem is viewed as a 
problem of the ``effective theory ideology''\cite{Salvio:2014soa}.

In view of the enormous success that these ideas have gained in the last years\cite{Meissner:2006zh,Salvio:2014soa,Boyle:2011fq,Alexander-Nunneley:2010tyr,Carone:2013wla,Farzinnia:2013pga,Ghilencea:2015mza,Ghilencea:2016dsl,Guo:2014bha,Kawamura:2013kua,Foot:2007iy,Meissner:2007xv,Oda:2018zth,Ghilencea:2016ckm,Heikinheimo:2013fta,Mooij:2018hew,Shaposhnikov:2008xi,Bezrukov:2014ipa,Bezrukov:2007ep,Bars:2013yba,Farina:2013mla,Brivio:2017vri,Steele:2013fka,Wang:2015cda}, it is of the greatest importance to investigate on these issues.  
Is it really possible that DR encodes {\it physical} properties that makes it the correct tool to calculate the quantum fluctuations in QFT, while the Wilsonian strategy produces unphysical terms? 

If this would turn out to be the case, such a finding would represent a breakthrough, and the physical mechanism that makes $m_H^2 \ll \Lambda^2$ would be uncovered. In this respect, it is worth to stress that the more common lore is that, although DR is a powerful technique to calculate radiative corrections, it has no direct physical interpretation.

One of the main goals of the present work is to perform a thorough analysis of the Wilsonian and DR methods for calculations in QFT.  
From this analysis the {\it physical meaning} and the {\it limits} of DR will clearly emerge. In our opinion this represents a relevant progress in our understanding of renormalization, that allows to make a correct use of DR.
Notable recent examples of physical effects that cannot be captured by DR calculations are in \cite{DeAlwis:2021gou,deAlwis:2021zab}, where it is shown that an effective field theory can be derived from String Theory only if a Wilsonian perspective is adopted. In this framework the decoupling of
states above and below the physical cut-off scale can be derived (contrary to what happens when calculations are performed in DR), and, despite largely diffused Swampland arguments, this allows for a positive value for the cosmological constant at cosmological scales, even if a negative value is found at the string/Kaluza-Klein (KK) scale. 
An older example is given by theories with spontaneous symmetry breaking (SSB) where the RG flow of the coupling constants deviates significantly from the perturbative one when the infrared (IR) region is approached\cite{Alexandre:1997gj}, and eventually the RG equations get totally modified\cite{Alexandre:1998ts}. 

With regard to the naturalness/hierarchy problem, another largely considered possibility consists in assuming that the UV completion of the SM provides for  $m_H^2(\Lambda)$ a value $m_H^2(\Lambda) \ll \Lambda^2$. In this scenario, once the problem of the \vv large correction'' is fixed in the UV, the RG equation for the running $m_H^2(\mu)$ provides the \vv small'' measured value  for the physical mass $m_H^2(v)$, which turns out to be of the same order of $m^2_H(\Lambda)$; then it seems that the naturalness/hierarchy problem can be solved this way. From this standpoint, the \vv explanation'' for the smallness of $m^2_H(v)$ is pushed towards the unknown realm of the Standard Model UV completion (quantum gravity, string theory, ...): the higher energy theory should operate the \vv miracle'' of leaving us with a value of $m_H^2(\Lambda)$ much lower than the naturally expected one, $m_H^2(\Lambda) \sim \Lambda^2$. 

Somehow complementary to the previous one, another scenario considers that at the scale $\Lambda$ (as naturally expected) $m_H^2(\Lambda) \sim \Lambda^2$. A much lower value of $m_H^2$ is obtained at the Fermi scale, through a mechanism presented  as \vv self-organized criticality'', where
the critical regime $m_H^2(v) \ll m_H^2(\Lambda)$ should be reached without resorting to any fine-tuning. 

In the present work we carefully  investigate both these scenarios, and show that our previous results on the physical meaning of the renormalization procedures are crucial to ascertain the viability of these proposed mechanisms.
Although it is not immediately apparent, we will see that they both hide a fine-tuning that makes them unfit to solve the naturalness/hierarchy problem.

The rest of the paper is organized as follows. To pave the way to our analysis, in Section 2 we briefly review the main steps that lead to the renormalized one-loop effective potential of a single component scalar field theory in $d=4$ dimensions. We use dimensional regularization, momentum cut-off, and  Wilsonian flow equations, and compare the results. In Section 3 we calculate the effective potential of a scalar theory in $d$-dimensions with the help of a momentum cut-off, establish the connection between WEFT strategy and DR calculations, and show how DR hides the necessary fine-tuning. In Section 4 we consider the Wilsonian RG flows, showing again how the usual perturbative RG equations hide the fine-tuning. In Section 5 we apply the results of the previous sections to the SM, and compare our results with previous literature. Section 6 is for the conclusions. 

\section{Dimensional regularization and WEFT}

In the last years there has been growing consensus  on the idea that DR might play a special role in defining QFTs, and that the DR and the WEFT approaches are physically different, with the latter missing some \vv truth'' that is instead  encoded in DR\cite{Meissner:2006zh,Salvio:2014soa,Boyle:2011fq,Alexander-Nunneley:2010tyr,Carone:2013wla,Farzinnia:2013pga,Ghilencea:2015mza,Ghilencea:2016dsl,Guo:2014bha,Kawamura:2013kua,Foot:2007iy,Meissner:2007xv,Oda:2018zth,Ghilencea:2016ckm,Heikinheimo:2013fta,Mooij:2018hew,Shaposhnikov:2008xi,Bezrukov:2014ipa,Bezrukov:2007ep,Bars:2013yba,Farina:2013mla,Brivio:2017vri,Steele:2013fka,Wang:2015cda}.  From this perspective,  WEFT is viewed as a calculation procedure that {\it improperly insists} in defining the renormalized theory through the successive elimination of  modes,
and the naturalness problem  as an {\it artefact} of this approach\cite{Salvio:2014soa}.

Clearly, if one would find that DR is really endowed with special properties that make it the {\it  correct physical way} to define QFTs, while the WEFT paradigm has to be discarded, this would be an  earthquake for our understanding of QFTs, and the naturalness/hierarchy problem would simply evaporate.  

The present section is devoted to a thorough investigation of these issues. For our analysis, 
a great simplification comes from the observation that, in order to investigate upon these questions, there is no need to consider the full SM. As we will see, the essence of the problem is entirely captured by considering the simpler $\phi^4$ theory. Moreover, it will be sufficient to restrict ourselves to the lowest order of approximation, focusing in particular on the one-loop effective potential $V(\phi)$. 
We stress that, when calculated with the help of a momentum cut-off, the one-loop potential provides an implementation (and an approximation) of the WEFT strategy (see Section 2.3). 

For a single component $d$-dimensional $\phi^4$ theory, the action is
\be\label{lagra}
S[\phi]=\int\,d^d\,x\left(\frac12\partial_\mu\phi\,\partial^\mu\phi +\Omega_0+\frac{1}{2}m_0^2\phi^2 +\frac{\mu^{4-d}\lambda_0}{4!}\phi^4 \right),
\ee
where $\mu$ is a mass scale introduced to keep $\lambda_0$ dimensionless, 
\begin{eqnarray}\label{counter}
\Omega_0 =\Omega+\delta \Omega \quad , \quad
m^2_0=m^2+\delta m^2 \quad , \quad
\lambda_0 = \lambda+\delta\lambda\,,
\end{eqnarray}
are the bare parameters, \,$\delta \Omega$, \,$\delta m^2$, and \,$\delta \lambda$ \,the counterterms,  $\Omega$, $m^2$ and $\lambda$ the renormalized parameters. The one-loop effective potential is
\begin{eqnarray}\label{effpotdimreg-c}
V_{1l}\left(\phi\right)= \Omega_0 + \frac{m_0^2}{2}\phi^2+
\frac{\mu^{4-d}\lambda_0}{4!}\phi^4 + \frac{1}{2}\int\frac{d^dk}{(2\pi)^d}
\ln\left(1+\frac{m^2 + \frac{1}{2}\mu^{4-d}\lambda\,\phi^2}{k^2}\right).
\end{eqnarray}

The integral in (\ref{effpotdimreg-c}) converges only for $d<2$, that for integer values of $d$ means only for  $d=1$, while for $d \geq 2$ it is UV divergent. 
Integrating over the angular variables, and defining
\begin{equation}
M^2(\phi) \equiv m^2+\frac{1}{2}\mu^{4-d}\lambda\,\phi^2,
\end{equation}
for the one-loop correction to the potential $\delta V(\phi)$ we have
\begin{eqnarray}\label{two}
\delta V=\frac{1}{2}\int\frac{d^dk}{(2\pi)^d}
\ln\left(1+\frac{M^2(\phi)}{k^2}\right) = 
\frac{1}{2(4\pi)^{\frac d2}\Gamma\left(\frac d2\right)}  \int_0^\infty{dk^2}\,(k^2)^{\frac d2 - 1}\,
\ln\left(1+\frac{M^2(\phi)}{k^2}\right).
\end{eqnarray}

We now briefly review the steps that lead to the renormalized one-loop potential, using first DR and then a momentum cut-off $\Lambda$ to calculate (\ref{two}) (where the latter, as stressed above, realizes an approximation of the WEFT strategy). We reassure the experts, 
familiar with 
the few lines of Sections 2.1 and 2.2,
that they are reported here only as functional to our analysis.

\subsection{Effective potential in $\boldsymbol{4d}$. Dimensional regularization}
The first observation that leads to the DR strategy consists in noting that the right hand side of \eqref{two} can be extended to complex $d$, but converges only for $\mathrm{Re}(d)<2$. We have 
\begin{align}\label{changevar} 
&\int_0^\infty\,{dk^2}\,(k^2)^{d/2 - 1}\,\ln\left(1+\frac{M^2(\phi)}{k^2}\right) 
=\frac 2 d \, \mu^d\,\left(\frac{M^2(\phi)}{\mu^2}\right)^{\frac d2}\,B\left(1-\frac d2,\frac{d}{2}\right)\nn \\
&=\frac 2 d \,\mu^d\, \left(\frac{M^2(\phi)}{\mu^2}\right)^{\frac d2}\,\Gamma \left(1-\frac d2\right) \, \Gamma\left(\frac d2 \right) \,,
\end{align}
where $B(\alpha,\beta)$ and $\Gamma(z)$ are the special Beta and Gamma functions. Together with some of their properties, they are given in Appendix A.

The second observation is that, if we replace $\Gamma(z)$ with its analytic extension $\overline \Gamma(z)$ (see Appendix A), the second line of (\ref{changevar})
can be extended to generic complex values of $d$. The function $\overline \Gamma(z)$ is obtained with the help of the Weierstrass representation for  $\Gamma(z)^{-1}$ (see \eqref{Weier}), from which we see that $\overline{\Gamma}(z)$ has poles for $z=0,-1,-2,\dots$

Inserting the last member of (\ref{changevar}) (with $\Gamma(z)$ replaced by $\overline \Gamma(z)$) in the right hand side of (\ref{two}), and using the relation 
$\overline \Gamma(z+1)=z\,\overline \Gamma(z)$, the DR rules for calculating $\delta V$ are as follows. First we replace (for any complex $d\neq 2,4,6, ...$)
\be \label{betaintfinal}
\frac{1}{2}\int\frac{d^dk}{(2\pi)^d}
\ln\left(1+\frac{M^2(\phi)}{k^2}\right) \to -\frac{\mu^d}{2(4\pi)^{\frac d2}}\,\left(\frac{M^2(\phi)}{\mu^2}\right)^{\frac d2}\,\overline \Gamma\left(-\frac d2\right).
\ee
Successively we expand the right hand side of (\ref{betaintfinal}) around $d=4$ ($\epsilon\equiv 4-d$), 
\begin{align}
\label{result dimreg}
\frac 1 2\int\frac{d^dk}{(2\pi)^d}
\ln\left(1+\frac{M^2(\phi)}{k^2}\right)\equiv \frac{\mu^{-\epsilon}\left[M^2(\phi)\right]^2}{64 \pi^2} \left( -\frac 2 \epsilon +\gamma +\ln \frac{M^2(\phi)}{4 \pi \mu^2}-\frac 32 \right)+\mathcal O (\epsilon) \,.
\end{align}
Then we cancel the pole in $\epsilon$ with the help of the couterterms $\delta \Omega$,  $\delta m^2$, $\delta \lambda$ in (\ref{counter}),
that in the $\overline{MS}$ scheme are $\left(\overline \epsilon\equiv\epsilon \left(1+\frac{\epsilon}{2}\ln \frac{e^{\gamma}}{4\pi}\right)\right)$
\begin{align}\label{lastlastct}
\delta \Omega =\frac{m^4}{32\pi^2 \overline \epsilon}\mu^{-\epsilon} \quad , \quad
\delta m^2=\frac{\lambda m^2}{16\pi^2\overline\epsilon} \quad , \quad
\delta \lambda=\frac{3 \lambda^2}{16\pi^2\overline\epsilon}\,.
\end{align}
Finally we take the limit $\epsilon \to 0$, and the renormalized one-loop potential reads (for $\Omega=0$)
\begin{eqnarray}\label{dimregc2}
V_{1l}(\phi)=\frac{1}{2}m^2\phi^2+\frac{\lambda}{4!}\phi^4
+\frac{1}{64\pi^2}\left(m^2+
\frac{\lambda}{2}\phi^2\right)^2\left[{\rm ln}
\left(\frac{m^2+\frac{\lambda}{2}\phi^2}{\mu^2}\right)
-\frac{3}{2}\right].
\end{eqnarray}

\subsection{Effective potential in $\boldsymbol{4d}$. Momentum cut-off}
If we calculate the loop integral in (\ref{effpotdimreg-c}) for $d=4$ with a sharp momentum cut-off, the one-loop correction to the potential is
\begin{align}\label{effec} 
\delta V (\phi)= 
\frac{1}{64\pi^2}\Biggl[\Lambda^4\,{\ln}
\left(1+\frac{M^2(\phi)}{\Lambda^2}
\right)+\Lambda^2 M^2(\phi)  
-\left[M^2(\phi)\right]^2{\ln}
\left(\frac{\Lambda^2+ M^2(\phi)}
{M^2(\phi)}\right)\Biggr] \,.
\end{align}
Taking $\frac{\phi^2}{\Lambda^2}$, $\frac{m^2}{\Lambda^2} \ll 1$,
and expanding the right hand side of (\ref{effec}) in powers of $\frac{M^2}{\Lambda^2}$:
\begin{align}\label{newVf}
V_{1l} (\phi)=\Omega_0+\frac{m_0^2}{2}\phi^2+\frac{\lambda_0}{4!}\phi^4
+
\frac{\Lambda^2 M^2}{32\pi^2}
-\frac{\left(M^2\right)^2}{64\pi^2} \left({\ln}
\frac{\Lambda^2}{ M^2} +\frac12\right)+ O\left(\frac{\phi^6}{\Lambda^2}\right).
\end{align}
Inserting (\ref{counter}) in\,(\ref{newVf}), with 
\bea\label{delta2}
\delta \Omega &=&-\frac{\Lambda^2 m^2}{32\pi^2}+\frac{m^4}{64\pi^2}\left[
{\rm ln\left(\frac{\Lambda^2}{\mu^2}\right)} 
-1 \right] \,; \,
\delta m^2 = -\frac{\lambda \Lambda^2}{32\pi^2}
+\frac{\lambda m^2}{32\pi^2}\left[
{\rm ln\left(\frac{\Lambda^2}{\mu^2}\right)} 
-1 \right] \,  ;
\nn\\
\delta \lambda&=&\frac{3\lambda^2}{32\pi^2}\left[
{\rm ln\left(\frac{\Lambda^2}{\mu^2}\right)} 
-1\right] \,,
\eea
and neglecting the cut-off suppressed terms\, $\frac{\phi^6}{\Lambda^2}$, $\frac{\phi^8}{\Lambda^4}$, ..., for the renormalized one-loop potential we find the same result obtained with DR, i.e.\,Eq.\,(\ref{dimregc2}). 

\subsection{Wilsonian RG flow and one-loop effective potential}

As mentioned above, the one-loop effective potential calculated with a momentum cut-off provides an approximation to the potential obtained within the WEFT framework. To elucidate this point, let us consider
the Wilsonian effective action $S_k[\phi]$, where $k$ is the running scale. Given the bare (tree-level) action $S_\Lambda[\Phi]$ (where $\Phi(x)= \sum_{0<|p|<\Lambda} \varphi_{_p} e^{ipx}$), $S_k[\phi]$ is obtained decomposing $\Phi(x)= \phi(x) + \phi'(x)$ (with
$\phi(x)= \sum_{0<|p|<k} \varphi_{_p} e^{ipx}$ and $\phi'(x)= \sum_{k<|p|<\Lambda} \varphi_{_p} e^{ipx}$),
and  integrating out the modes $\varphi_{_p}$ in the range $k<p<\Lambda$:
\begin{equation}\label{wils1}
e^{-S_{k}[\phi]}\equiv\int D[\phi']e^{-S_{\Lambda}[\phi+\phi']}. 
\end{equation}
The effective action is $\Gamma[\phi]= S_{k=0}[\phi]$, while the bare action is $S_\Lambda[\phi]=S_{k=\Lambda}[\phi]$.

At the infinitesimally lower scale $k-\delta k$, the Wilsonian action $S_{k-\delta k}[\phi]$ is obtained through an equation similar to (\ref{wils1})
\begin{equation}\label{wils2}
e^{-S_{k-\delta k}[\phi]}=\int D[\phi']e^{-S_{k}[\phi+\phi']}
\end{equation}
where $\phi'$ contains only modes in the infinitesimal shell \, $k-\delta k<p<k$.

Let us consider the gradient expansion for $S_k[\phi]$,  
\begin{equation} \label{wilsonact}
S_k[\phi]=\int\,d^dx\left(U_k(\phi) +\frac{Z_k(\phi)}{2}  \,\partial_\mu\phi\,\partial_\mu\phi 
+Y_k(\phi)\,(\partial_\mu\phi\,\partial_\mu\phi)^2 +W_k(\phi)\,(\phi\,\partial_\mu\partial_\mu\phi)^2 + \cdots \right),
\end{equation}
and restrict ourselves to the Local Potential Approximation (LPA), that amounts to
\begin{equation}
Z_k(\phi)=1\quad, \quad Y_k(\phi)= W_k(\phi) = \cdots=0\,.
\end{equation}

By taking as background field $\phi(x)$ the homogeneous configuration
\be
\phi(x)=\phi_0\,,
\ee
and performing in \eqref{wils2} the integration over $\phi'$ under the assumption that the saddle point $\phi'_{\rm sp}$ is trivial\footnote{The modifications to Eq.\,\eqref{wils3} that arise when a non-trivial saddle point $\phi'_{\rm sp} \neq 0$ is present are discussed in\cite{Alexandre:1998ts}.}, i.e. $\phi'_{\rm sp}=0$, we get ($U''_k(\Phi)\equiv\frac{\partial^2 U_k(\Phi)}{\partial\Phi^2}$)

\begin{equation}\label{wils3}
U_{k-\delta k}(\phi_0)=  
U_k(\phi_0)+ \frac{1}{2}\int'\frac{d^d p}{(2\pi)^d}\ln\left(\frac{p^2+U''_k(\phi_0)}{p^2}\right)
\end{equation}
where the prime indicates that the integration is performed within the shell $[k-\delta k, k]$, and we have subtracted a field independent term.  
In the limit $\delta k \to 0$, we finally have

\begin{equation}\label{Ukrge}
k\frac{\partial}{\partial k }U_k(\phi_0)=-\frac{k^d}{(4\pi)^{\frac d2}\Gamma \left(\frac d2\right)}\ln\left(\frac{k^2+U''_k(\phi_0)}{k^2}\right),
\end{equation}
that is the RG flow equation for the Wilsonian potential $U_k(\phi_0)$ in the LPA.

This is an intrinsically non-perturbative equation for $U_k(\phi)$, that 
implements the WEFT strategy. However, its non-perturbative nature becomes evident only for sufficiently small values of $k$, the IR regime (that will be better specified in Section 4). On the contrary,
for sufficiently large values of $k$ (UV regime), it reproduces the perturbative results.  

As mentioned above, for $k=0$ the Wilsonian potential $U_k(\phi)$ is the effective potential $V_{\rm eff}(\phi)$, while for $k=\Lambda$ it is the bare (tree-level) potential $U_\Lambda(\phi)$. 
We now show under which approximation the perturbative one-loop effective potential $V_{1l}(\phi)$ is obtained from \eqref{Ukrge}. Taking for $U_\Lambda(\phi)$
\be\label{barepot}
U_\Lambda(\phi)= \Omega_0 + \frac{m_0^2}{2}\phi^2+
\frac{\mu^{4-d}\lambda_0}{4!}\phi^4\,,
\ee
and approximating $U_k(\phi)$ in the right hand side of (\ref{Ukrge})   with $U_\Lambda(\phi)$ (i.e.\,freezing $U_k(\phi)$ to its boundary value at $k=\Lambda$), we can integrate both sides of (\ref{Ukrge}) in the whole momentum range $[0,\Lambda]$ (indicated by the upper case $(\Lambda)$ in the integral below), and get 
\be\label{effectpotenz}
V_{1l}(\phi)= \Omega_0 + \frac{m_0^2}{2}\phi^2+
\frac{\mu^{4-d}\lambda_0}{4!}\phi^4 + \frac{1}{2}\int^{(\Lambda)}\frac{d^dk}{(2\pi)^d}
\ln\left(1+\frac{m_0^2 + \frac{1}{2}\mu^{4-d}\lambda_0\,\phi^2}{k^2}\right),
\ee 
that is nothing but the one-loop effective potential  (\ref{effpotdimreg-c}) once we replace 
the bare values $m_0^2$ and $\lambda_0$ in the above integral with the corresponding renormalized values, which is coherent with the fact that the one-loop correction is $O(\hbar)$. 

Eq.\,\eqref{effectpotenz} shows that the one-loop potential calculated with the hard cut-off $\Lambda$ provides a specific implementation (and approximation) of the WEFT strategy. A smoothed equivalent implementation of WEFT is obtained by means of the proper-time regularization, and in Appendix B we give an example of that.    
From now on, we will refer to the WEFT strategy having in mind one-loop calculations of the kind considered in this section.

\subsection{DR versus WEFT}
Let us compare now the DR and WEFT approaches for the calculation of the one-loop effective potential.  
Apart from the elementary observation that 
the two procedures give the same result  (once the counterterms are appropriately chosen), we would like to make a couple of other  comments, relevant to our subsequent analysis. 

From the results briefly reviewed above, it seems that 
DR intrinsically 
avoids: 
\begin{itemize}
\item the appearance of quadratic divergences, so that there is no need to fine-tune the scalar mass (the same holds true for the cosmological constant);  
\item the appearance of higher powers\, $\phi^6$, $\phi^8$, ...,  with coupling constants of inverse mass power dimensions, that on the contrary are present in WEFT (see (\ref{newVf})). 
\end{itemize}
Moreover, when $m^2$ vanishes, at the quantum level the theory (that is clearly scale invariant at the classical level) shows only a soft (logarithmic) breaking of scale invariance, and the scalar mass is generated through the CW 
mechanism\cite{Meissner:2006zh,Salvio:2014soa,Boyle:2011fq,Alexander-Nunneley:2010tyr,Carone:2013wla,Farzinnia:2013pga,Ghilencea:2015mza,Ghilencea:2016dsl,Guo:2014bha,Foot:2007iy,Meissner:2007xv,Oda:2018zth,Ghilencea:2016ckm,Heikinheimo:2013fta,Mooij:2018hew,Shaposhnikov:2008xi,Steele:2013fka,Wang:2015cda}.

Accordingly,  some authors speculate that DR might grasp an element of truth that is missed by the WEFT scheme. 
Even more, the physical essence of the WEFT approach is questioned. The whole idea of including in the theory the quantum fluctuations via iterative integrations over infinitesimal momentum shells is considered as misleading, and the naturalness/hierarchy problem is regarded as an artefact of \vv the Effective Field Theory ideology"\cite{Salvio:2014soa}. 

We will show that, contrary to these expectations, DR does not encode any special physical principle, but is a {\it specific way} of implementing the WEFT paradigm that can be applied only when the perturbative expansion is valid.
In particular, we will see that 
DR realizes the fine-tuning of the mass parameter, although it does it in a {\it hidden} way. 
To this end, in the next section we turn our attention to the calculation of the one-loop effective potential in $d$ dimensions by means of a momentum cut-off.  

\section{Effective potential in d dimensions}
\label{edsc}
As a first step of our analysis, we calculate the loop integral in\,\eqref{effpotdimreg-c}, that is the one-loop correction $\delta V(\phi)$ for the $d$-dimensional theory, by introducing a cut-off $\Lambda$ ($u\equiv\frac{M^2}{M^2+\Lambda^2}$)
\begin{align}\label{int}
&\delta V(\phi)=\frac{1}{2}\int^{(\Lambda)}\frac{d^dk}{(2\pi)^d}
\ln\left(1+\frac{M^2}{k^2}\right)  
\nn\\
&=	\frac{\mu^d}{d(4\pi)^{\frac d2}\Gamma\left( \frac d2\right)} \left[\, 
\left(\frac{M^2}{\mu^2}\right)^{d\over2}\int_{u}^{1} dt \,(1-t)^{{d\over2}-1}\,t^{-{d\over2}}
+\left(\frac{\Lambda}{\mu}\right)^{d} \ln \left ( 1+ \frac{M^2}{\Lambda^2} \right )\, \right]\nonumber\\
&\equiv \delta V_1(\phi) + \delta V_2(\phi)
\end{align}
where we defined 
\begin{align}
\label{dV1}
\delta V_1(\phi)&=\frac{ \mu^d}{d(4\pi)^{\frac d2}\Gamma\left( \frac d2\right)}
\left(\frac{M^2}{\mu^2}\right)^{d\over2}\int_{u}^{1} dt \,(1-t)^{{d\over2}-1}\,t^{-{d\over2}},\\
\label{dV2}
\delta V_2(\phi)&=\frac{\mu^d}{d(4\pi)^{\frac d2}\Gamma\left( \frac d2\right)} \left(\frac{\Lambda}{\mu}\right)^{d} \ln \left ( 1+ \frac{M^2}{\Lambda^2} \right ) .
\end{align}

It is not difficult to show that, for any value of the dimension $d$ (we should not forget that $d$ is a positive integer), we can write  
\begin{equation}\label{newsplitting0}
\int_{u}^{1} dt \,(1-t)^{{d\over2}-1}\,t^{-{d\over2}}
=\lim_{z \to d} \left[\overline B\left(1-\frac{z}{2},\frac{z}{2}\right) - \overline B_i\left(1-\frac{z}{2},\frac{z}{2};u\right)\right],
\end{equation}
where $\overline B(\alpha,\beta)$ and $\overline{B}_i(\alpha,\beta;x)$ are the analytic extensions of the complete and incomplete Beta functions $B(\alpha,\beta)$ and  $B_i(\alpha,\beta;x)$, and are  defined for $x\in \mathbb{R}$ and for generic complex values of $\alpha$ and $\beta$, excluding $\alpha,\beta= 0,-1,-2,\dots$ (the functions $B$, $B_i$, $\overline{B}$, $\overline{B}_i$, with the corresponding existence domains and some of their properties, are given in Appendix A).

From Appendix A we know that 
\begin{eqnarray}\label{Bbar.p}
\overline B\left(1-\frac{z}{2},\frac{z}{2}\right)=\overline \Gamma\left(1-\frac{z}{2}\right) \overline \Gamma\left(\frac{z}{2}\right),
\end{eqnarray}
which shows that $\overline B\left(1-\frac{z}{2},\frac{z}{2}\right)$  has poles in $z=2,4,6,\dots$ (see (\ref{Weier})). Moreover,  expanding $\overline B_i\left(1-\frac{z}{2},\frac{z}{2}, u \right)$ in powers of ${M^2}/{\Lambda^2}\ll1$ 
(remember that \,$u=\frac{M^2}{M^2+\Lambda^2}$),
we have
\begin{eqnarray}\label{betaiexp}
\overline B_i\left(1-\frac{z}{2},\frac{z}{2}, u\right)=\frac{2}{2-z}
\left(\frac{M^2}{\Lambda^2}\right)^{\frac{2-z}{2}}
-\frac{2}{4-z}\left(\frac{M^2}{\Lambda^2}\right)^{\frac{4-z}{2}}+\frac{2}{6-z}\left(\frac{M^2}{\Lambda^2}\right)^{\frac{6-z}{2}}
+\dots\,,
\end{eqnarray}
which shows that $\overline B_i\left(1-\frac{z}{2},\frac{z}{2}, u \right)$ has the same poles as
$\overline B\left(1-\frac{z}{2},\frac{z}{2}\right)$.

Finally, for each of the values $z=2,4,6, \dots$, we can show that the poles developed in (\ref{Bbar.p}) cancel those coming from (\ref{betaiexp}). For definiteness, in what follows we consider in (\ref{newsplitting0}) only the case $d=4$, i.e. the theory defined in $d=4$ dimensions,
but the calculations and considerations developed below can be repeated for any of the values $d=2,4,6,\dots$ 

Let us go back to (\ref{int}), the one-loop correction $\delta V(\phi)=\delta V_1(\phi)+\delta V_2(\phi) $ to the potential for the  $d$-dimensional theory. As $\Lambda$ is finite, the second member of this equation is finite for any integer value of $d$ (actually it is finite for any complex value of $d$, with ${\rm Re}\, d>0$). From \eqref{dV1} and (\ref{newsplitting0}) we have: 
\begin{eqnarray}\label{loopintegralA}
\delta V_1(\phi) =\frac{\mu^d}{d(4\pi)^{\frac d2}\Gamma\left( \frac d2\right)}
\left(\frac{M^2}{\mu^2}\right)^{d\over2}\int^1_{u}dt\,
t^{-\frac{d}{2}}(1-t)^{\frac{d}{2}-1}\equiv \lim_{z\to d}\left[A_1(z)-A_2(z)\right],
\end{eqnarray}
where we defined
\begin{eqnarray}\label{A1}
A_1(z)&\equiv& F(z)\cdot\overline B \left(1-\frac{z}{2},\frac{z}{2}\right) \\
\label{A2}
A_2(z)&\equiv& F(z)\cdot\overline B_i \left(1-\frac{z}{2},\frac{z}{2};u\right)\\
\label{A3}
{\rm with} \,\,\,\,\,\,\,\ F(z)&\equiv& \frac{\mu^z}{z(4\pi)^{\frac z2}\Gamma\left( \frac z2\right)}
\left(\frac{M^2}{\mu^2}\right)^{z\over2}.
\end{eqnarray}

With the help of (\ref{Bbar.p}) and (\ref{betaiexp}), we can expand $F$, $\overline B$ and $\overline B_i$ around $z=4$. More specifically,  we write $z=4-\epsilon$ and expand these functions around $\epsilon=0$, thus getting 
\begin{align}
\label{effe}
&F(4-\epsilon)=\frac{\mu^{-\epsilon}}{64 \pi^2}\,\left[M^2(\phi)\right]^2 \left[1+\left(-\gamma_{_{E}}+\log(4\pi)-\log\frac{M^2(\phi)}{\mu^2}+\frac 32\right)\frac{\epsilon}{2}\right] + O(\epsilon^2)\\
\label{bbarra}
&\overline B \left(-1+\frac{\epsilon}{2},2-\frac{\epsilon}{2}\right) = -\frac{2}{\epsilon} + O(\epsilon)\\
\label{bbarrai}
&\overline B_i \left(-1+\frac{\epsilon}{2},2-\frac{\epsilon}{2},\frac{M^2}{M^2+\Lambda^2}\right) = -\frac{2}{\epsilon}-\frac{\Lambda^2}{M^2}+\log \frac{\Lambda^2}{M^2}  + O(\epsilon).
\end{align}
Then, using (\ref{effe}), (\ref{bbarra}) and (\ref{bbarrai}), we can  write $A_1(4-\epsilon)$ and $A_2(4-\epsilon)$ as:
\begin{align}\label{epsA1}
A_1(4-\epsilon)&=\frac{\mu^{-\epsilon}\left[M^2(\phi)\right]^2}{64 \pi^2} \left( -\frac 2 \epsilon +\gamma +\ln \frac{M^2(\phi)}{4 \pi \mu^2}-\frac 32 \right)+\mathcal{O}(\epsilon)\\
\label{epsA2}
A_2(4-\epsilon)
&=-\frac{\mu^{-\epsilon}}{64 \pi^2}\,\left[M^2(\phi)\right]^2 \left(\frac{\Lambda^2}{M^2(\phi)}-\log\frac{\Lambda^2}{M^2(\phi)}\right) \nn \\
& + \frac{\mu^{-\epsilon}\left[M^2(\phi)\right]^2}{64 \pi^2} \left( -\frac 2 \epsilon +\gamma +\ln \frac{M^2(\phi)}{4 \pi \mu^2}-\frac 32 \right) +\mathcal{O}(\epsilon)
+\mathcal{O}\left(\frac{M^2}{\Lambda^2}\right)\,.
\end{align}

Let us make now two observations that are crucial to our analysis.
From (\ref{epsA1}) and (\ref{epsA2}) we see that in the difference $A_1 (4-\epsilon)- A_2 (4-\epsilon)$, that is nothing but  Eq.\,(\ref{loopintegralA}) for  $z=4-\epsilon$, the polar terms $\frac{1}{\epsilon}$ cancel each other (as expected), and the limit $\epsilon \to 0$ can be safely and easily taken. For $d=4$ we 
have
\begin{align}\label{dV1cut}
\delta V_1 (\phi) \Big |_{d=4} &=\lim_{\epsilon \to 0} \left[ A_1(4-\epsilon) - A_2(4-\epsilon)\right] \nn \\
&=-\frac{1}{64 \pi^2}\,\left[M^2(\phi)\right]^2 \left(\frac{\Lambda^2}{M^2(\phi)}-\log\frac{\Lambda^2}{M^2(\phi)}\right)+\mathcal{O}\left(\frac{1}{\Lambda^2} \right) .
\end{align}
Taking now the surface term $\delta V_2(\phi)$ in \eqref{dV2} for the case $d=4$, performing the expansion in $M^2/\Lambda^2\ll1$, and combining the result with $\delta V_1(\phi)$ above, we finally get
\begin{align}\label{newcutpot}
\delta V (\phi)=\delta V_1+ \delta V_2=
\frac{\Lambda^2 M^2(\phi)}{32\pi^2}
-\frac{\left[M^2(\phi)\right]^2}{64\pi^2} \left({\ln}
\frac{\Lambda^2}{ M^2(\phi)} +\frac12\right)+ O\left(\frac{\phi^6}{\Lambda^2}\right),
\end{align}
that is nothing but (\ref{newVf}), i.e.\,\,the result 
obtained directly for the 4-dimensional theory when the loop integral is calculated  with a cut-off. The quadratic and logarithmic divergences in  (\ref{newcutpot}) are then cancelled with the help of the counterterms (\ref{delta2}), and this finally gives the renormalized potential (\ref{dimregc2}). 

Our second observation is that (keeping aside for a moment the $\epsilon\to 0$ limit) if we conveniently write
the one-loop correction to the potential as
\begin{align}
\label{Vtotale}
\delta V(\phi)=\delta V_1(\phi)+ \delta V_2(\phi)=A_1(4-\epsilon)\, +\,\left[\delta V_2(\phi)-A_2(4-\epsilon)\right],
\end{align}
and neglect the term in the square bracket of the last member (with no justification for the moment; we will comment on this point below), 
we have
\be
\delta V(\phi)\label{V1ldr} =
A_1(4-\epsilon)=\frac{\mu^{-\epsilon}\left[M^2(\phi)\right]^2}{64 \pi^2} \left( -\frac 2 \epsilon +\gamma +\ln \frac{M^2(\phi)}{4 \pi \mu^2}-\frac 32 \right)+\mathcal{O}(\epsilon)\,,
\ee
that is nothing but the DR result (\ref{result dimreg}) for $\delta V$. Taking the counterterms (\ref{lastlastct}), for the renormalized potential $V_{1l}(\phi)$ we again obtain (\ref{dimregc2}).

Referring to (\ref{int}), we proceed with our analysis by noting that $\delta V$ in $d=4$ dimensions can be calculated  in one of the following three equivalent ways: 
\vskip 4pt
(a) taking $d= 4$, and then calculating the integral; 
\vskip 4pt
(b) calculating the integral for generic $d$, and then replacing $d=4$; 
\vskip 4pt
(c) calculating separately $A_1(4-\epsilon)$ and $A_2(4-\epsilon)$, expanding each of them around 

\noindent
\,\,\,\,\,\,\,\,\,\,\,\,\,\,\,\, $\epsilon=0$, considering the difference $A_1-A_2$, and finally taking the limit $\epsilon \to 0$.
\vskip 4pt
The procedure (c) is the one that we used in this section, and is certainly more intricate, and definitely much longer and cumbersome than (a) and/or (b). However, for the purposes of our analysis, that is to \textit{uncover the physical meaning} of DR, we need to refer to this one. 

We have just seen that
if we neglect\, $A_2$ \,and\, $\delta V_2$ \,in\,  \eqref{Vtotale}, we are left with the DR result.
But what could justify the neglect  of\, $A_2$\, and\, $\delta V_2$\, in (\ref{Vtotale})?
To answer this question, we begin by noting that, 
when we use the procedure (c), we can write
\be
V_{1l}(\phi)= \label{V1ltot}
\lim_{\epsilon \to 0}\left[\Omega + \delta \Omega +\frac12 (m^2+ \delta m^2 )\,\phi^2+\frac{\mu^\epsilon}{4!}(\lambda +
 \delta \lambda)\,\phi^4 + 
A_1(4-\epsilon)-A_2(4-\epsilon)\, +\,\delta V_2(\phi)\right], 
\ee   
where $\delta V_2(\phi)$ is given by (\ref{dV2}) (with $d$ replaced by $4-\epsilon$), while $A_1(4-\epsilon)$ and $A_2(4-\epsilon)$ are given by (\ref{epsA1}) and (\ref{epsA2}) respectively. As in the difference $A_1(4-\epsilon) - A_2(4-\epsilon)$ the polar terms in $\epsilon$ disappear, 
in (\ref{V1ltot}) we are left with the original WEFT result, and the divergences in $\Lambda$  are cancelled by the counterterms $\delta\Omega$, $\delta m^2$ and $\delta \lambda$ given in (\ref{delta2}).

However, we now follow a different pattern, that {\it naturally} leads to the DR recipes, and allows to find the {\it physical meaning} of DR. As we will see, this represents an advancement in our understanding of renormalization, that allows to avoid misinterpretations and misuses of DR.

Going back to the splitting (\ref{Vtotale}) for $\delta V$, and defining $\Delta V_2(\phi)$ as given below,
\begin{align}\label{DV2}
&\delta V(\phi)= 
A_1(4-\epsilon)\, +\,\left[\delta V_2(\phi)-A_2(4-\epsilon)\right]
\equiv A_1(4-\epsilon)\, + \Delta V_2(\phi)  ,
\end{align}
our objective is to realize the  cancellation of the divergences separately in $A_1(4-\epsilon)$ and in $\Delta V_2(\phi)$,
starting with $\Delta V_2(\phi)$. Note that, while in $A_1(4-\epsilon)$ only divergences for $\epsilon \to 0$ appear, $\Delta V_2(\phi)$ contains divergences for $\epsilon \to 0$ as well as for $\Lambda \to \infty$.

In order to realize such a separate cancellation, we begin by making the splitting
\be
\delta \Omega = \delta \Omega_1 + \delta \Omega_2 \quad , \quad 
\delta m^2 = \delta m^2_1 + \delta m^2_2 \quad , \quad 
\delta \lambda = \delta \lambda_1 + \delta \lambda_2 .
\ee
Choosing
\bea\label{delta22}
\delta \Omega_2 &=&-\frac{\Lambda^2 m^2}{32\pi^2}+\frac{m^4}{64\pi^2}\left[
{\rm ln\left(\frac{\Lambda^2}{\mu^2}\right)} -1 \right] 
-\frac{m^4}{32\pi^2 \overline \epsilon}\mu^{-\epsilon}\\
\label{delta23}
\delta m^2_2 &=& -\frac{\lambda \Lambda^2}{32\pi^2}
+\frac{\lambda m^2}{32\pi^2}\left[
{\rm ln\left(\frac{\Lambda^2}{\mu^2}\right)} -1 \right] -\frac{\lambda m^2}{16\pi^2\overline\epsilon} \\
\label{delta24}
\delta \lambda_2&=&\frac{3\lambda^2}{32\pi^2}\left[
{\rm ln\left(\frac{\Lambda^2}{\mu^2}\right)}-1\right] - \frac{3 \lambda^2}{16\pi^2\overline\epsilon},
\eea
and inserting (\ref{delta22}), 
(\ref{delta23}), and (\ref{delta24}) in (\ref{V1ltot}), we have 
\be \label{V1ltot2}
V_{1l}(\phi)=\lim_{\epsilon \to 0} 
\left[\Omega + \delta \Omega_1 +\frac{m^2+ \delta m_1^2}{2} \,\phi^2+\frac{\lambda + \delta \lambda_1}{4!}\,\phi^4 +  A_1(4-\epsilon)+\mathcal{O}\left(\frac{1}{\Lambda^2}\right)+\mathcal{O}(\epsilon)\right].
\ee 

Apart from the harmless $\mathcal{O}\left(\frac{1}{\Lambda^2}\right)$ and $\mathcal{O}(\epsilon)$ terms, (\ref{V1ltot2}) is the one-loop potential that we would have obtained following directly the DR \vv rules''.
In fact, the final form (\ref{dimregc2}) for the renormalized potential is obtained from (\ref{V1ltot2}) once $\delta \Omega_1$, $\delta m_1^2$, and $\delta \lambda_1$ are chosen according to the $\overline{MS}$ counterterms in (\ref{lastlastct}).

This is the result that allows to uncover the {\it physical content} of DR, and will enable us to answer one of the questions that motivated this work, namely whether or not DR is endowed with special properties that make it the {\it correct physical way} to define QFTs\cite{Salvio:2014soa}, thus helping in solving the naturalness problem.  

We have just shown that (\ref{V1ltot2}), which in a DR setting is obtained from the well-known recipes, actually comes from the introduction of an intermediate step in the process of obtaining the renormalized potential, that in the Wilsonian language (theory of critical phenomena) corresponds to the tuning toward the critical regime (critical surface). As already stressed (see Section 2.3), the calculation of the one-loop effective potential with a momentum cut-off $\Lambda$ is a specific implementation of the Wilson's strategy in the perturbative regime.  

The rules of DR are nothing but a \vv short-cut" that allows to derive for $V_{1l}(\phi)$ directly the right hand side of (\ref{V1ltot2}). We stress that, following the alternative (and longer) path (c), we have learnt that we are  never dispensed  from the {\it necessity} of subtracting the quadratically divergent contribution to the mass of the scalar particle. Such a subtraction is realized through the counterterm $\delta m_2^2$ in (\ref{delta24}). When we adopt the short-cut that takes (\ref{V1ltot2}) as the starting point for the calculation of the one-loop effective potential, this cancellation is {\it hidden}, and it seems we are dispensed of it. 

\begin{figure}[t]
	\centering
	\begin{tikzpicture}[baseline=-3]
	\begin{feynman}
	\vertex[draw, circle, minimum size=1cm,very thick] (m1) at (-4, 0) {$1$};
	\vertex[label={center:{\small Bare Potential}}] (m10) at (-4,-1) {};
	\vertex[draw, circle, minimum size=1cm,very thick] (m2) at (4, 0) {$3$};
	\vertex[label={center:{\small Renormalized Potential}}] (m20) at (4,-1) {};
	\vertex[draw, circle, minimum size=1cm,very thick, label={\small DR Bare Potential}] (m3) at (0, 3.5) {$2$};
	\diagram* {
		(m1) -- [fermion,edge label'={\small $\delta \Omega,\,\delta m^2,\,\delta \lambda$}] (m2);
		(m1) -- [fermion,half left, out=50, in=140,edge label={\small $\delta \Omega_{2},\,\delta m^2_{2},\,\delta \lambda_{2}$}] (m3);
		(m3) -- [fermion,half left, out=40, in=130,edge label={\small $\delta \Omega_{1},\,\delta m^2_{1},\,\delta \lambda_{1}$}] (m2);
	};
	\end{feynman}
	\end{tikzpicture}
	\caption{Bubble \vv $1$" represents the bare potential \eqref{barepot},
		bubble \vv $3$" the renormalized potential \eqref{dimregc2}. 
		The line connecting \vv $1$" with \vv $3$" represents the calculation of the one-loop potential in $d=4$ dimensions with a UV cut-off, and the  determination of the counterterms \eqref{delta2}. Bubble \vv $2$" represents the bare potential \eqref{V1ltot2} in DR language. The line connecting \vv $1$" with \vv $2$" represents the calculation of the one-loop potential performed in $d$-dimensions with a UV cut-off, and the determination of the counterterms \eqref{delta22}-\eqref{delta24}. The line connecting \vv $2$" with \vv $3$" represents the determination of the $\overline{MS}$ counterterms \eqref{lastlastct}.}
	\label{Stepfigure}
\end{figure}
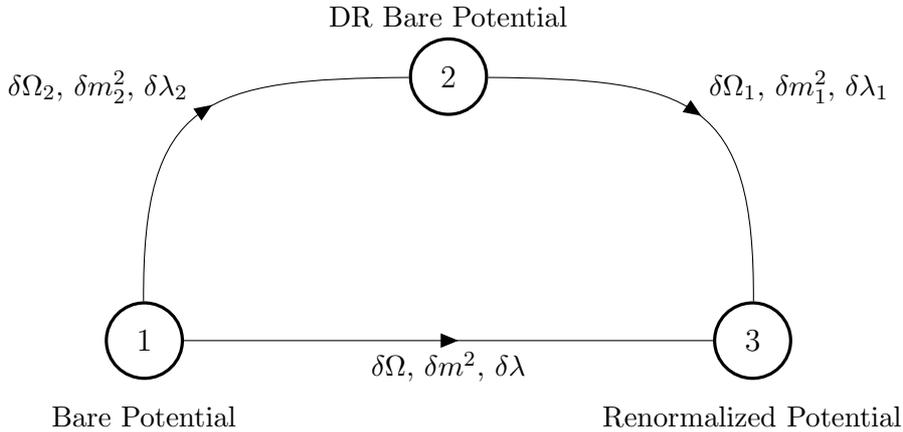

The above remarks are illustrated in Fig.\,\ref{Stepfigure}. Let us start by considering the \vv Bare Potential", represented by bubble 
$
\begin{tikzpicture}[baseline=-3]
\begin{feynman}
\vertex[draw, circle, minimum size=0.5cm,very thick] (m) at (0, 0) {\footnotesize{$1$}};
\diagram* {
	(m)
};
\end{feynman}
\end{tikzpicture}
$\,,
and defined by $V(\phi)=\Omega_0+\frac{1}{2}m_0^2\phi^2+\frac{\lambda_0\mu^{4-d}}{4!}\phi^4$. 
To obtain the one-loop correction $\delta V=\delta V_1+\delta V_2$, we {\it have to} sum (integration in $d^dk$) over the momenta of the intermediate virtual states, see \eqref{int}, \eqref{dV1} and \eqref{dV2}. The explicit calculation of this sum, arrested to the maximal value $|k|=\Lambda$, allows to determine the counterterms \,$\delta \Omega$, $\delta m^2$ and $\delta \lambda$, so that the one-loop \vv Renormalized Potential"  $V_{1l}(\phi)=V(\phi)+\delta V(\phi)$ of bubble 
$
\begin{tikzpicture}[baseline=-3]
\begin{feynman}
\vertex[draw, circle, minimum size=0.5cm,very thick] (m) at (0, 0) {\footnotesize{$3$}};
\diagram* {
	(m)
};
\end{feynman}
\end{tikzpicture}
$
is obtained. In Fig.\,\ref{Stepfigure} this calculation is represented  by the line that connects 
$
\begin{tikzpicture}[baseline=-3]
\begin{feynman}
\vertex[draw, circle, minimum size=0.5cm,very thick] (m) at (0, 0) {\footnotesize{$1$}};
\diagram* {
	(m)
};
\end{feynman}
\end{tikzpicture}
$
with 
$
\begin{tikzpicture}[baseline=-3]
\begin{feynman}
\vertex[draw, circle, minimum size=0.5cm,very thick] (m) at (0, 0) {\footnotesize{$3$}};
\diagram* {
	(m)
};
\end{feynman}
\end{tikzpicture}
$\,.

The same result is obtained by following a different, but {\it totally equivalent}, pattern. We can go to bubble 
$
\begin{tikzpicture}[baseline=-3]
\begin{feynman}
\vertex[draw, circle, minimum size=0.5cm,very thick] (m) at (0, 0) {\footnotesize{$3$}};
\diagram* {
	(m)
};
\end{feynman}
\end{tikzpicture}
$\,, 
the \vv Renormalized Potential", passing first from bubble 
$
\begin{tikzpicture}[baseline=-3]
\begin{feynman}
\vertex[draw, circle, minimum size=0.5cm,very thick] (m) at (0, 0) {\footnotesize{$2$}};
\diagram* {
	(m)
};
\end{feynman}
\end{tikzpicture}
$\,. 

The divergences in $\Lambda$ are {\it cancelled} 
in step\,
$
\begin{tikzpicture}[baseline=-3]
\begin{feynman}
\vertex[draw, circle, minimum size=0.5cm,very thick] (m) at (0, 0) {\footnotesize{$1$}};
\diagram* {
	(m)
};
\end{feynman}
\end{tikzpicture}
\to
\begin{tikzpicture}[baseline=-3]
\begin{feynman}
\vertex[draw, circle, minimum size=0.5cm,very thick] (m) at (0, 0) {\footnotesize{$2$}};
\diagram* {
	(m)
};
\end{feynman}
\end{tikzpicture}
$ (see \eqref{delta22}-\eqref{delta24}). This is the \textit{hidden fine-tuning} that leads to the \vv DR Bare Potential".  The DR counterterms $\delta \Omega_1$, $\delta m_1^2$ and $\delta \lambda_1$ are determined in step $
\begin{tikzpicture}[baseline=-3]
\begin{feynman}
\vertex[draw, circle, minimum size=0.5cm,very thick] (m) at (0, 0) {\footnotesize{$2$}};
\diagram* {
	(m)
};
\end{feynman}
\end{tikzpicture}
\to
\begin{tikzpicture}[baseline=-3]
\begin{feynman}
\vertex[draw, circle, minimum size=0.5cm,very thick] (m) at (0, 0) {\footnotesize{$3$}};
\diagram* {
	(m)
};
\end{feynman}
\end{tikzpicture}
$\,. The important lesson is that the cancellation of the quadratic divergences is {\it secretly} realized when, through the DR recipes, we directly access the \vv DR Bare Potential" of bubble
$
\begin{tikzpicture}[baseline=-3]
\begin{feynman}
\vertex[draw, circle, minimum size=0.5cm,very thick] (m) at (0, 0) {\footnotesize{$2$}};
\diagram* {
	(m)
};
\end{feynman}
\end{tikzpicture}
$.

The above results  show
that DR is a smart calculation technique that, when the conditions for the perturbative expansion are satisfied, implements both steps of the  {\it physical} Wilsonian EFT  calculation (the integration over the momentum modes, and the tuning towards the critical surface) at once\footnote{In Appendix B we consider a different but equivalent implementation of the WEFT strategy, using a proper-time regularization, and apply it to the two-point vertex function rather than to the full effective potential.}. In other words,  DR is an efficient technique that takes us {\it directly} to renormalized quantities, and, as such, is a very welcome tool. At the same time, our results show that DR 
is not endowed with any special physical property, despite claims to the contrary\cite{Meissner:2006zh,Salvio:2014soa} that received a large follow-up\cite{Boyle:2011fq,Alexander-Nunneley:2010tyr,Carone:2013wla,Farzinnia:2013pga,Ghilencea:2015mza,Ghilencea:2016dsl,Guo:2014bha,Kawamura:2013kua,Foot:2007iy,Meissner:2007xv,Oda:2018zth,Ghilencea:2016ckm,Heikinheimo:2013fta,Mooij:2018hew,Shaposhnikov:2008xi,Bezrukov:2014ipa,Bezrukov:2007ep,Farina:2013mla,Brivio:2017vri,Bars:2013yba,Wang:2015cda,Steele:2013fka}.

The above results can also be obtained by means of a formal short-cut, that again shows how the unwanted terms are secretly cancelled.

Let us start with (\ref{int}) for $\delta V=\delta V_1+\delta V_2$, where $\delta V_1$ and $\delta V_2$ are given in \eqref{dV1} and \eqref{dV2} respectively. Focusing on (\ref{dV1}), and relaxing the physical requirement that $d$ is an integer, we consider the integral in this equation for complex values of $d$, with $0<{\rm Re}\,d<2$. 
Under this hypothesis, we can operate the (mathematically legitimate) splitting 
\begin{align}\label{splitting2}
\int_{u}^{1} dt \,(1-t)^{{d\over2}-1}\,t^{-{d\over2}}
&= \int_{0}^{1} dt \,(1-t)^{{d\over2}-1}\,t^{-{d\over2}}
-\int_{0}^{u} dt \,(1-t)^{{d\over2}-1}\,t^{-{d\over2}},
\end{align}
and note that (remember that $u=\frac{M^2}{M^2+\Lambda^2}$) in (\ref{splitting2}) we can safely take the limit $\Lambda \to \infty$. Under this limit, the second term in the right hand side of \eqref{splitting2} vanishes, and we are left with the first term only, that is nothing but the Beta function $B(\alpha,\beta)$ (see Appendix A) of arguments $\alpha=1-\frac{d}{2}$ and $\beta=\frac d 2$. Moreover, going to (\ref{dV2}) for $\delta V_2$, we see that for $0<{\rm Re}\,d < 2$, we have $\lim_{\Lambda \to \infty}\delta V_2=0$. Therefore, under these conditions, $\delta V=\delta V_1$.
Replacing $B\left(1-\frac{d}{2},\frac{d}{2}\right)$ with its analytic extension $\overline{B}$, and pretending that we can extend the above results (obtained after sending $\Lambda \to \infty$\,!) to generic values of $d$, from (\ref{dV1}) we immediately get the DR result. 

The bottom line of the above derivation is that, operating with $0<{\rm Re}\,d < 2$, we can safely send the cut-off $\Lambda$ to infinity. Therefore, when we extend to higher dimensions the results obtained this way, we may get the wrong impression that there is no need for a physical cut in the sum over the loop momenta. 

\vskip 5pt

Finally, to better understand the physical meaning of DR, and the reason why it gives the correct result for the perturbatively renormalized quantities, let us further compare the detailed calculations for the one-loop potential within the Wilsonian and DR frameworks.

Let us begin by considering $A_2(4-\epsilon)$ in (\ref{epsA2}), and note that the terms $\log \Lambda^2/M^2$ (first line) and $2/\epsilon$ (second line) both come from the expansion around $\epsilon=0$ of the same term in (\ref{betaiexp}), namely the one with the pole in $d=4$. This is why they are multiplied by the same factor. 
On the other hand, $A_1(4-\epsilon)$ in (\ref{epsA1}) is given by the product of $F(4-\epsilon)$ times $\overline{B}\left(-1+\frac{\epsilon}{2},2-\frac{\epsilon}{2}\right)$ (see \eqref{A1}). A simple inspection of (\ref{effe}) and \eqref{bbarra} shows that the terms $2/\epsilon$ and $\log M^2/\mu^2$ in $A_1(4-\epsilon)$ have the same coefficient.
Moreover, we already noted that the $2/\epsilon$ polar terms in $A_1(4-\epsilon)$ and $A_2(4-\epsilon)$ {\it must} have the same coefficient, otherwise there would be no cancellation of these \vv spurious singularities". Therefore, $\log M^2/\mu^2$ in $A_1$, and $\log M^2/\Lambda^2$ in $A_2$ {\it must} have the same coefficient.

This latter observation proves that the renormalized potential obtained from the DR rules {\it must} be the same as the renormalized potential derived from the physical Wilsonian calculation in the perturbative regime. In fact, the relevant part of the one-loop correction to the potential calculated with DR is the $\log M^2/\mu^2$ mentioned above, and contained in $A_1$. At the same time, the coefficient of the divergent term $\log \Lambda^2/M^2$, obtained when performing the Wilsonian calculation, is {\it the same} as the coefficient of the similar term in $A_2$ (see (16) and (43)). Therefore the WEFT calculation needs counterterms (see (\ref{delta2})) that add up to a $\log \Lambda^2/\mu^2$ with {\it exactly the same coefficient}.

This simple chain of observations shows why, under the condition of the perturbative regime, the DR formal rules provide for the effective potential (and more generally for any physical quantity) {\it exactly the same result} that is obtained when the direct WEFT physical calculation is performed. We have actually shown that the DR rules are obtained in the WEFT framework, and are far from being in contrast with it. DR is a technique able to give the physically correct perturbative results, although the deep physical reason for that is very much hidden in the procedure.

\vskip 5pt

Before ending this section we would like to note that the results of our analysis are of particular interest when studying BSM models with classical scale invariance, where the use of dimensional regularization seems to suggest that scale invariance (apart from a welcome logarithmic violation) can be preserved also at the quantum level \cite{Salvio:2014soa,Meissner:2006zh,Boyle:2011fq,Alexander-Nunneley:2010tyr,Carone:2013wla,Farzinnia:2013pga,Ghilencea:2015mza,Ghilencea:2016dsl,Guo:2014bha,Kawamura:2013kua,Foot:2007iy,Meissner:2007xv,Oda:2018zth,Ghilencea:2016ckm,Heikinheimo:2013fta,Mooij:2018hew,Shaposhnikov:2008xi,Bezrukov:2014ipa,Bezrukov:2007ep,Bars:2013yba,Wang:2015cda,Steele:2013fka}. 
To better illustrate our point, let us refer in particular  to\cite{Salvio:2014soa,Meissner:2006zh}. By considering 
the possibility that the fundamental theory of nature does not possess any mass or length scale,  in \cite{Salvio:2014soa} only dimension four operators are kept, more precisely SM operators with dimensionless couplings, with the Higgs field non-minimally coupled to gravity. 
Similarly, in \cite{Meissner:2006zh} a conformally extended version of the Standard Model is considered, with right-chiral neutrinos and a minimally
enlarged scalar sector. In both cases, it seems that with these almost scale invariant models the naturalness/hierarchy problem is absent.

However, the reason why we only see a logarithmic violation of scale invariance is entirely due to the fact that the quantum corrections are calculated with DR. This is why the particle masses, generated through the Coleman-Weinberg mechanism, seem to not exhibit strong UV sensitivity.
It is obvious that, if we consider a theory that contains only dimension four operators, and at the same time compute the radiative corrections with DR, operators of dimension two can never be generated, so that we get the impression that no fine-tuning is needed. However, as we have shown in the present section, DR contains a \vv hidden fine-tuning", and the fact that dimension two operators do not appear when calculating radiative corrections is simply due to that. In this respect, we note that in the original Coleman and Weinberg calculation the effective potential is obtained introducing a momentum cut-off for the loop integrals, and the renormalized $m^2=0$ mass is obtained only after performing the fine-tuning \cite{Coleman:1973jx}.

\section{Perturbative, Wilsonian, and subtracted RG}
Let us consider the Callan-Symanzik equation for the renormalized potential of Eq.\,(\ref{dimregc2}), obtained by requiring independence of $V_{1l}(\phi)$ from $\mu$
\begin{equation}\label{Seq}
\mu\,\frac{d}{d \mu}V_{1l}(\phi)=\left(\mu\, \frac{\partial}{\partial \mu}+\beta_{_\Omega}\,\frac{\partial}{\partial \Omega}+{m^2}\gamma_{_m}\,\frac{\partial}{\partial m^2}+\beta_{_\lambda}\,\frac{\partial}{\partial \lambda}\right)V_{1l}(\phi)=0\,.
\end{equation}
Inserting (\ref{dimregc2}) in (\ref{Seq}) the one-loop RG functions read
\begin{eqnarray}
\label{betamu1}
\beta_{_\Omega}&=& \mu\frac{d\Omega}{d\mu}=\frac{ m^4}{32\pi^2} \\
\label{betamu2}
\gamma_{_m}&=&\frac{1}{m^2}\left(\mu\frac{d m^2}{d\mu}\right)=\frac{ \lambda}{16\pi^2} \\
\label{betamu3}
\beta_{_\lambda}&=& \mu\frac{d\lambda}{d\mu}=\frac{3 \lambda^2}{16\pi^2}\,.
\end{eqnarray} 

Below we will see that, once the subtraction that leads to the tuning towards the critical surface is performed, these RG functions coincide with the corresponding Wilsonian ones in the UV region (perturbative regime). We note that they are obtained in the perturbative regime, and can be derived either using DR or the momentum cut-off calculation.  

The flow of the bare parameters is obtained in a similar way, requiring independence of the bare effective potential in \eqref{newVf} from $\Lambda$. From an equation analogous to \eqref{Seq}, we obtain
\begin{eqnarray}\label{baref1}
\Lambda\frac{d}{d\Lambda}\Omega_0&=&-\frac{m_0^2 \Lambda^2}{16\pi^2}+\frac{ m_0^4}{32\pi^2} \\
\label{baref2}
\Lambda\frac{d}{d\Lambda}m_0^2&=&-\frac{\lambda_0 \Lambda^2}{16\pi^2}+\frac{\lambda_0 m_0^2}{16\pi^2}\\
\label{baref3}
\Lambda\frac{d}{d\Lambda}\lambda_0&=&\frac{3\lambda_0^2}{16\pi^2}  \,,
\end{eqnarray}
that, as we show below, are nothing but the Wilsonian RG equations for the running parameters in the UV regime. 

To ascertain this point, let us go back to Eq. (\ref{Ukrge}) for the Wilsonian potential $U_k(\phi)$, that for the reader's convenience we write here for $d=4$
\begin{equation}\label{Ukrge2}
k\frac{\partial}{\partial k }U_k(\phi)=-\frac{ k^4}{16\pi^2}\ln\left(\frac{k^2+U''_k(\phi)}{k^2}\right).
\end{equation}
Inserting in (\ref{Ukrge2}) the expansion
\begin{equation}\label{Uk}
U_k(\phi)=\Omega_k+\frac{1}{2} m^2_{k} \phi^2+\frac{1}{4!}\lambda_k \phi^4+\frac{1}{6!}\lambda_k^{(6)}\phi^6+\frac{1}{8!}\lambda_k^{(8)}\phi^8+\dots\,,
\end{equation}
where the Wilsonian RG parameters are (the upper label $(i)$ denotes the $i$-th derivative with respect to $\phi$)
\begin{equation}
\Omega_k=U_k(0) \,\, , \,\, m_k^2=U_k^{(2)}(0)\,\, , \,\, \lambda_k=U_k^{(4)}(0)\,\,, \,\, \lambda^{(6)}_k=U_k^{(6)}(0)\,\,, \,\, \lambda^{(8)}_k=U_k^{(8)}(0)\,\,,\,\, \dots
\end{equation}
for $\Omega_k$, $m^2_k$, $\lambda_k$, ... we easily get

\begin{align}
\label{omegak}
k \frac{\partial \Omega_k}{\partial k}&=-\frac{k^4}{16\pi^2}\log \left(\frac{k^2+m^2_k}{k^2}\right) \\
\label{mk}
k \frac{\partial m^2_k}{\partial k}&=-\frac{k^4}{16\pi^2}\frac{\lambda_k}{k^2+m^2_k} \\
\label{lambdak}
k \frac{\partial \lambda_k}{\partial k}&=-\frac{k^4}{16\pi^2}\left(\frac{\lambda^{(6)}_k}{k^2+m^2_k}-\frac{3 \lambda_k^2}{(k^2+m_k^2)^2}\right) \\
k \frac{\partial \lambda^{(6)}_k}{\partial k}&=-\frac{k^4}{16\pi^2}\left(\frac{\lambda^{(8)}_k}{k^2+m^2_k}-\frac{15 \lambda_k\,\lambda_k^{(6)}}{(k^2+m_k^2)^2}+\frac{30 \lambda_k^3}{(k^2+m^2_k)^3}\right) \\
\dots \nn
\end{align}

These are the Wilsonian renormalization group equations in the framework of the local potential approximation. They form a set of infinitely many coupled differential equations, and govern the non-perturbative flow of the theory parameters. If\, $k^2+m^2_k >0$\, in the whole range $[0,\Lambda]$, this flow essentially coincides with the perturbative one (see below). Similar results can be obtained for theories with scalars and fermions\cite{Clark:1992jr,Krajewski:2014vea}.

If, on the contrary, there exists a critical value $k_{\rm cr}$ where $k_{\rm cr}^2+m^2_{k_{\rm cr}}=0$, that is the case when the theory manifests SSB, the non-perturbative nature of these equations becomes manifest when the region\, $k_{\rm cr}^2+m^2_{k_{\rm cr}}\gtrsim 0$\, is approached. In this regime, the flow of the coupling constants deviates significantly from the perturbative one\cite{Alexandre:1997gj}. For values of $k<k_{\rm cr}$, that is within the spinodal instability region, the flow equation \eqref{Ukrge2} no longer holds, and has to be replaced with a new RG equation, that realizes the Maxwell construction for the SSB potential\cite{Alexandre:1998ts}.

Limiting ourselves to the case when\, $k^2+m^2_k >0$, and retaining for the potential $U_k(\phi)$ in (\ref{Uk}) only terms up to the quartic coupling $\lambda_k$, this set of equations is truncated to Eqs.\,(\ref{omegak}), (\ref{mk}) and (\ref{lambdak}) only, where in the latter the term with $\lambda_k^{(6)}$ is missing. 

Under the condition $k^2 \gg m^2_k$, i.e. in the UV regime, expanding these three equations in $m_k^2/k^2$, we easily get
\begin{eqnarray}
\label{UVomegak}
k\frac{\partial}{\partial k}\Omega_k&=&-\frac{k^2m_k^2 }{16\pi^2}+\frac{m_k^4}{32\pi^2} \\
\label{UVmk}
k\frac{\partial}{\partial k}m_k^2&=&-\frac{k^2 \lambda_k}{16\pi^2}+\frac{\lambda_k m_k^2}{16\pi^2}\\
\label{UVlambdak}
k\frac{\partial}{\partial k} \lambda_k&=&\frac{3 \lambda_k^2}{16\pi^2}  \,,
\end{eqnarray}
that coincide with \eqref{baref1}, \eqref{baref2} and \eqref{baref3} (once we replace $k$ with $\Lambda$), that is what we wanted to show.

To understand the relation between the renormalized flow (Eqs.\,\eqref{betamu1}, \eqref{betamu2} and \eqref{betamu3}) and the Wilsonian one (Eqs.\,\eqref{baref1}, \eqref{baref2} and \eqref{baref3}, or equivalently \eqref{UVomegak}, \eqref{UVmk} and \eqref{UVlambdak}), we have to introduce first two \vv critical" parameters. Let us start with the mass. From the finite difference version of \eqref{baref2} (or equivalently \eqref{UVmk}), we have
\begin{equation}\label{finitediff}
m_0^2\left(\Lambda-\delta\Lambda\right)=m_0^2\left(\Lambda\right)+\frac{\delta \Lambda}{\Lambda}\,\frac{\lambda_0\left(\Lambda\right)}{16\pi^2}\,\Lambda^2-\frac{\delta \Lambda}{\Lambda}\,\frac{\lambda_0\left(\Lambda\right)m_0^2\left(\Lambda\right)}{16\pi^2}+\mathcal O\left(\frac{\delta \Lambda^2}{\Lambda^2}\right).
\end{equation}

We now define the subtracted mass parameter\, $\widetilde m^2(\Lambda-\delta \Lambda)$\, at the scale\, $\Lambda-\delta \Lambda$\, through the equation
\begin{align}\label{sub1}
\widetilde{m}^2(\Lambda-\delta \Lambda)&\equiv m_0^2(\Lambda-\delta \Lambda)-m^2_{\rm cr}(\Lambda)\,,
\end{align}
where
\begin{equation}
m^2_{\rm cr}(\Lambda)\equiv\frac{\lambda_0\left(\Lambda\right)}{16\pi^2}\,\Lambda\, \delta \Lambda
\end{equation}
is the {\it critical mass} that comes from the integration in the momentum shell $[\Lambda-\delta \Lambda,\Lambda]$, and  vanishes in the $\delta \Lambda\to 0$ limit, so that we have the boundary
\begin{align}\label{sub2}
\widetilde{m}^2(\Lambda)&= m_0^2(\Lambda).
\end{align}
With the help of (\ref{sub1}) and (\ref{sub2}), Eq.\,(\ref{finitediff}) can be written (in differential form) as
\begin{equation}
\label{msub}
\frac{1}{\widetilde m^2}\left(\Lambda\frac{d}{d\Lambda}\widetilde m^2\right)=\frac{\lambda_0}{16\pi^2}.
\end{equation}

Comparing \eqref{msub} with \eqref{betamu2}, we see that the perturbative flow of the renormalized mass $m^2$ is nothing but the flow of $\widetilde m^2$. The right hand side of \eqref{msub} is precisely the perturbative $\gamma_{_m}$ that appears in (\ref{betamu2}).

Similarly, by considering the finite difference version of \eqref{baref1} (or equivalently (\ref{UVomegak})), and
defining the subtracted vacuum energy $\widetilde \Omega$ through the equation
\begin{equation}
\widetilde \Omega(\Lambda-\delta \Lambda)\equiv\Omega_0(\Lambda-\delta \Lambda)-\Omega_{\rm cr}(\Lambda)\,,
\end{equation}
where
\begin{equation}
\Omega_{\rm cr}(\Lambda)\equiv\frac{\widetilde m^2(\Lambda) }{16\pi^2}\,\Lambda\,\delta \Lambda\,,
\end{equation}
is the {\it critical vacuum energy},
for $\widetilde \Omega$ we obtain the flow equation
\begin{eqnarray}\label{Omegasub}
\Lambda\frac{d}{d\Lambda}\widetilde\Omega=\frac{\widetilde m^4}{32\pi^2} \,.
\end{eqnarray}

As before, comparing \eqref{Omegasub} with  (\ref{betamu1}), we see that the perturbative flow of the renormalized vacuum energy $\Omega$ coincides with the flow of $\widetilde \Omega$, and the right hand side of \eqref{Omegasub} is nothing but the perturbative $\beta_{_\Omega}$ of \eqref{betamu1}.

For the dimensionless coupling $\lambda$ there is obviously no subtraction to operate, and in fact comparing \eqref{baref3} with (\ref{betamu3}) we immediately see that the perturbative flow equation for the renormalized coupling $\lambda$ coincides with the UV flow of $\lambda_0$.

For the purposes of our analysis it is important to stress that the perturbative flow equations of the positive mass dimension parameters $m^2$ and $\Omega$, that can be obtained by using either DR or a momentum cut-off, are nothing but the RG equations of the {\it fine-tuned} parameters $\widetilde{\Omega}$ and $\widetilde m^2$ in the UV regime, i.e.\,the (UV) flow of the Wilsonian parameters subtracted of their critical values. This corresponds to the tuning towards the critical surface. We have then shown that the renormalized RG equations \eqref{betamu1}, \eqref{betamu2} and \eqref{betamu3} contain the fine-tuning.

In the next section we apply to the Standard Model the results and considerations developed here.

\section{Standard Model. Perturbative and Wilsonian RG}

The fact that at LHC no new particles have been observed allows to speculate that the SM could be valid all the way up to the Planck scale $M_P$, or to another high energy scale as for instance $\Lambda_{\rm GUT}$, or even a transplanckian scale. From now on we indicate this ultimate UV scale with $\Lambda$. For energies above this scale, we can imagine different scenarios: the SM could be replaced by a different theory, outside the QFT paradigm (string theory, loop quantum gravity, ...), or it might even be that, merging with quantum gravity, it could be extrapolated up to infinitely large energies\cite{Shaposhnikov:2009pv}.  

\vskip 5pt

Starting with the appropriate boundary conditions at $\Lambda$, the RG flows that connect $\Lambda$ to the Fermi scale $\mu_F$ \,should provide the measured values of the coupling constants and of the particle masses as RG outputs at the scale $\mu_F$. Let us concentrate on the running of the Higgs boson mass $m^2_H(\mu)$. 
As mentioned in the Introduction, a boundary condition typically regarded as a possible solution to the naturalness/hierarchy problem\cite{Giudice:2013yca,Holthausen:2013ota} is the so-called \vv miracle"
\begin{equation}
\label{Little mass}
m^2_H(\Lambda)\ll \Lambda^2,	
\end{equation} 
that could come as a left-over of the Standard Model UV completion. 

One specific implementation of \eqref{Little mass} is obtained imposing the Veltman condition, i.e.\,the vanishing of the quadratic divergences\cite{Veltman:1980mj}. In the pure SM such a condition is verified at $\Lambda \sim 10^{23}$ GeV (when the couplings are run with two-loop RG functions)\cite{Hamada:2012bp,Jones:2013aua}. If we would like to implement the Veltman condition at the Planck scale, $\Lambda=M_P$, we should consider extensions of the SM, as for instance in\cite{Chankowski:2014fva}. For the purposes of our analysis, however, it is totally irrelevant whether we consider the SM or a modified version of it. For this reason, in the following we concentrate on the SM.

Let us consider the perturbative RG flows, and restrict ourselves to the two-loop approximation for the RG functions (see for instance\cite{Ford:1992mv},\cite{Holthausen:2011aa})
\begin{align}
\label{perturbative RG SM}
\mu \frac{d}{d \mu} \lambda_i&= \beta_{_{\lambda_i}}\,, \\
\label{runmassren}
\mu \frac{d}{d \mu} m_H^2&=m_H^2\gamma_{_m} \,,
\end{align}
where $\lambda_i$ ($i=1, \dots, 5$) stands for the SM quartic coupling $\lambda$, the top Yukawa coupling $y_t$ and the three gauge couplings $g_i$.

When $\gamma_{_m}$ takes on perturbative values, i.e.\,\,\,$\gamma_{_m} \ll 1$ (which is the case in the SM), and the RG equation for $m_H^2(\mu)$ is given by \eqref{runmassren}, we certainly have $m^2_H(\Lambda)\sim m^2_H(\mu_F)$. For instance, choosing $\mu_F=m_t$ and taking for $m_H(m_t)$ the value $m_H=125.7$ GeV, if we take for $\Lambda$ the scale where the Veltman condition is satisfied, Eq.\,(\ref{runmassren}) imposes the boundary $m_H(\Lambda)=129.87$ GeV.
More generally, similar results are obtained whenever the UV condition \eqref{Little mass} is satisfied.
Therefore, when the (\ref{Little mass}) miracle is realized, and the RG mass flow is governed by \eqref{runmassren}, it seems that the naturalness/hierarchy problem is solved.

However we note that the miracle
(\ref{Little mass}) can effectively protect $m_H^2$ from large quantum corrections only if the SM really provides the multiplicative renormalization encoded in (\ref{runmassren}). 
This latter condition is necessary to obtain $m^2_H(m_t)\sim m^2_H(\Lambda)$, i.e. the absence of hierarchy.  
From the previous Section we know that what runs in \eqref{runmassren} is not the original Wilsonian mass $m^2_H(\mu)$, but rather the subtracted (i.e.\,renormalized) Higgs mass $\widetilde m_H^2(\mu)$, where the fine-tuning of the quadratic divergence is already performed (see (\ref{msub})).
Therefore, we cannot couple Eqs.\,\eqref{Little mass} and \eqref{runmassren} and pretend that the result $m^2_H(m_t) \sim m_H^2(\Lambda)$ provides a solution to the naturalness/hierarchy problem. In fact, whatever boundary $m_H^2(\Lambda)$ we use (including the boundary (\ref{Little mass})), if we do not subtract the critical value of the mass,
nothing can protect $m_H^2$ from getting a \vv quadratically divergent" ($\sim \Lambda^2$) contribution. As stressed in the previous Section, such a subtraction is nothing but the {\it fine-tuning}, and is necessary to switch from bare to renormalized mass. 

It is worth to stress again that 
the SM is an EFT, where  the physical UV cut-off $\Lambda$ plays the role of a distinguished scale, above which its UV completion has to be considered. 
But physics below $\Lambda$ is governed by the SM, so the only consistent way of getting physical quantities is through effective quantum field theory calculations. Therefore, the appearance of quadratically divergent (i.e. proportional to $\Lambda^2$) contributions to the mass cannot be avoided: to reach the Fermi scale value $m_H \sim 125.7$\,GeV, a fine-tuning {\it must} be operated.

Moreover we note that, as correctly pointed out in \cite{DeAlwis:2021gou,deAlwis:2021zab}, to derive from string theory (or from any other UV completion) an effective field theory, it is necessary to take a Wilsonian perspective. In these recent papers it was shown that the decoupling of states above and below the KK scale (i.e. the cut-off scale for the \vv low energy'' effective theory) in a type IIB string theory
can be demonstrated only by means of a bona fide Wilsonian calculation. Applying that to the well-known problem concerning the sign of the cosmological constant (CC), the author shows that at cosmological scales a positive value for CC can be obtained even if a negative value is found at the KK scale.

\vskip 5pt

Other attempts to solve the naturalness/hierarchy problem, either within the SM, or in the framework of some SM extension, are based on the RG equations for the coupling constants of interest (the quartic Higgs coupling and/or additional couplings when SM extension are considered), with boundary conditions again fixed in the UV\cite{Chankowski:2014fva,Holthausen:2011aa,Iso:2012jn,Ibanez:2013gf,Masina:2013wja,Hashimoto:2013hta,Hashimoto:2014ela,Haba:2013lga}. Running the quartic coupling $\lambda$ down to the Fermi scale, the Higgs mass $m_H$ is determined through the usual relation between $\lambda$ and the vacuum expectation value (vev) of the Higgs field\,\footnote{According to the chosen renormalization conditions, this relation is either the tree-level or the radiatively corrected one\cite{Sirlin:1985ux}.}. It might seem that in this manner the problem of quadratic divergences is avoided. However, we should not forget that the vev gets a radiative correction from tadpole diagrams that, if not cancelled, bring $\sim \Lambda^2$ contributions to the mass, and we have the same fine-tuning problem considered above. Moreover, although in some of these works\cite{Chankowski:2014fva,Holthausen:2011aa,Iso:2012jn,Hashimoto:2013hta,Hashimoto:2014ela} a sort of (softly broken) conformal invariance is apparently implemented, all these approaches still contain a hidden fine-tuning. The physical quantities are either obtained within a DR calculation, or by means of subtracted RG equations.  

\vskip 5pt
 
Differently from the hidden fine-tuning of the previously discussed approaches, in\cite{Wetterich:1981ir,Wetterich:1983bi,Wetterich:1990an,Wetterich:1991be,Bornholdt:1992up,Shaposhnikov:2009pv,Wetterich:2011aa,Wetterich:2016uxm,Pawlowski:2018ixd} the presence of quadratic divergences is properly acknowledged, and it is correctly pointed out that they locate the critical surface in the couplings space.
Moreover, the authors stress that the quantities we are interested in are the deviations of the physical parameters from their critical values (not the bare ones), that are nothing but the renormalized parameters. While this observation is correct, it does not give any indication on the physical mechanism that triggers the approach to the critical surface. Although the authors refer to their approach as to a \vv self-organized criticality" phenomenon, they explicitly perform the subtraction of the quadratically divergent term in the mass parameter. Without such a subtraction, the system would never be driven towards the critical surface. In other words, they perform the usual fine-tuning of the theory. In a true self-organized critical phenomenon, a dynamical mechanism drives the physical system towards the critical surface, and no subtraction has ever to be performed.

The observation that the bare couplings and the critical surface are not universal quantities, and that \vv usually in quantum field theories they are not of much interest", is certainly not a justification for performing the subtraction by hand. The essence of the naturalness problem consists in searching for the physical mechanism responsible for the suppression of the large radiative corrections to the mass. In \cite{Wetterich:1981ir,Wetterich:1983bi,Wetterich:1990an,Wetterich:1991be,Bornholdt:1992up,Shaposhnikov:2009pv,Wetterich:2011aa,Wetterich:2016uxm,Pawlowski:2018ixd} it is simply shown that, once the subtraction is performed by hand, with a large $\gamma_{_m}$ ($\sim 2$) the theory can accomodate a large hierarchy between the Fermi and the Planck scale. Moreover, the circumstance that different choices of the cut-off scheme can give different values for the coefficient of the quadratic divergences\cite{Pawlowski:2018ixd} is not a problem in itself. If we think, for instance, of a supersymmetric theory, the cancellation of the quadratic divergences is related to the simultaneous presence of bosonic and fermionic superpartners, and occurs whatever specific cut-off scheme is chosen.

\vskip 5pt

Finally, in a couple of recent papers\cite{Mooij:2021lbc,Mooij:2021ojy} it is suggested that there might exist formulations of QFT that \textit{ab initio} do not exhibit divergences. The authors claim that BPHZ is the most famous of these approaches (also more recent literature\cite{Lenshina:2020edt} is quoted). It is actually well known that, once in the BPHZ method the \vv R-operation" is applied, finite quantities are obtained. However, such an operation is related to the mathematical procedure introduced by Hadamard to extract the finite part of a divergent integral\cite{Hadamard1,Hadamard2}, and corresponds to the subtraction operated via counterterms.

The authors also claim that another notable example of finite formulation of QFT, based on the Callan-Symanzik equations\,\cite{Callan:1970yg,Symanzik:1970rt}, is presented in\cite{Blaer:1974foy,Callan:1975vs}. However, as clearly explained in Callan's lectures\cite{Callan:1975vs}, the renormalized finite results are obtained in two steps. First a modification of the functions in the loop integrals is obtained through a subtraction \vv à la Pauli-Villars", thus getting finite results. Successively, the renormalized quantities are obtained with the help of the Callan-Symanzik equations. Therefore, both BPHZ and the method explained in\cite{Callan:1975vs} are nothing else than implementations of the subtraction of  divergences, totally equivalent to the usual renormalization procedure, and as such cannot provide any {\it ab initio} finite formulation of QFT.

\section{Summary, conclusions and outlook}
This work is focused on the evergreen subject of renormalization, with particular reference to the SM and BSM theories, and more generally of theories containing scalar fields. We begin with a thorough analysis of dimensional regularization, usually considered as a useful calculation technique deprived of any direct physical interpretation. The analysis is done comparing the derivation of the one-loop effective potential in DR with a direct calculation performed in the framework of the Wilsonian effective field theory approach.

We have shown that DR implements at once both steps of the physical Wilsonian EFT calculation (the integration over the quantum fluctuation modes, and the tuning towards the critical surface), provided that the conditions for the validity of the perturbative expansion are fulfilled. As such, it is a practical and welcome tool. In Section 3 we have shown in detail that the DR results are nothing but an intermediate step in the Wilsonian derivation of renormalized physical quantities, where a {\it hidden} fine-tuning, that secretly realizes the tuning towards the critical region, is automatically encoded.

These findings enabled us to answer one of the physical questions that motivated the present study, a subject that has driven lot of recent research work\cite{Meissner:2006zh,Salvio:2014soa,Boyle:2011fq,Alexander-Nunneley:2010tyr,Carone:2013wla,Farzinnia:2013pga,Ghilencea:2015mza,Ghilencea:2016dsl,Guo:2014bha,Kawamura:2013kua,Foot:2007iy,Meissner:2007xv,Oda:2018zth,Ghilencea:2016ckm,Heikinheimo:2013fta,Mooij:2018hew,Shaposhnikov:2008xi,Bezrukov:2014ipa,Bezrukov:2007ep,Bars:2013yba,Farina:2013mla,Brivio:2017vri,Steele:2013fka,Wang:2015cda}, namely whether or not DR is endowed with special physical properties that make it the correct way to define QFTs. Our results definitely show that this is not the case. The physical mechanism that realizes the tuning toward the renormalized theory is not encoded in unknown physical properties of DR. As a consequence, DR cannot be of help in solving the naturalness/hierarchy problem. 

In particular, we have shown that BSM models based on (classically) scale-invariant extensions of the Standard Model, where (apart from weak violations) the scale invariance is kept also at the quantum level through the use of dimensional regularization, do not provide a solution to the naturalness/hierarchy problem, as hoped by their proponents. In fact, when the classical lagrangian contains only operators of dimension four, and dimensional regularization is used, no terms with dimension different than four can ever be generated. In particular, this is the case for operators of dimension two, and we could get the impression that no fine-tuning is needed, and that the naturalness/hierarchy problem could be solved this way. As we have shown, however, DR contains a \vv hidden fine-tuning", that invalidates such a conclusion.

We also analysed the naturalness/hierarchy problem in the renormalization group framework. According to recent literature, if the UV completion of the SM 
provides the boundary $m_H^2(\Lambda) \ll \Lambda^2$, the problem would disappear, as the perturbative anomalous dimension $\gamma_{_m}$ ($\ll 1$)
allows to get $m^2_H(\mu_F) \sim m^2_H(\Lambda)$\cite{Giudice:2013yca,Holthausen:2013ota}. In another scenario, where the same RG equation \eqref{runmassren} for $m^2_H(\mu)$ is used, it seems that the hierarchy $m^2_H(\mu_F) \ll m^2_H(\Lambda) \sim \Lambda^2$
can be accommodated, as the physical value $m_H^2(\mu_F)$ is obtained assuming that the UV completion of the SM provides for $\gamma_{_m}$ a non-perturbative value $\gamma_{_m} \gtrsim 2$\cite{Wetterich:1981ir,Wetterich:1983bi,Wetterich:1990an,Wetterich:1991be,Bornholdt:1992up,Shaposhnikov:2009pv,Wetterich:2011aa,Wetterich:2016uxm,Pawlowski:2018ixd}.
However, as shown in detail in Section 5,  the RG equation used for $m^2_H(\mu)$ already contains the fine-tuning, and in both these scenarios the suggested solution to the naturalness/hierarchy problem comes from this tuning. Therefore, they cannot solve the problem.

Again in the RG framework, the same question has been attacked in a different way, that might seem to circumvent the fine-tuning problem. Within the SM, or in one of its extensions, the RG equations for the couplings are considered with boundary conditions fixed in the UV\cite{Chankowski:2014fva,Holthausen:2011aa,Iso:2012jn,Ibanez:2013gf,Masina:2013wja,Hashimoto:2013hta,Hashimoto:2014ela,Haba:2013lga}. Running the Higgs quartic coupling $\lambda$ down to the Fermi scale, $m_H$ is determined through the usual relation between $\lambda$ and the vev of the Higgs field. However, we have shown that this does not solve the naturalness/hierarchy problem, as the vev gets radiative corrections from tadpole diagrams. If not cancelled (fine-tuning), they bring quadratic divergent contributions to $m^2_H$, again leaving the problem unsolved.

All these shortcomings are related in a way or another to the use of DR, of which we have shown the range of validity and limitations, providing examples where the direct physical Wilsonian calculations are needed, while DR gives incorrect results\cite{DeAlwis:2021gou,deAlwis:2021zab,Alexandre:1997gj,Alexandre:1998ts}.

We also analysed other recent attempts to solve the naturalness/hierarchy problem, that aim at a finite formulation of quantum field theories\cite{Mooij:2021lbc,Mooij:2021ojy}. We showed that these calculations actually implement the usual subtraction of  divergences, and therefore do not shade any light on the problem.

Before ending this section, we would like to discuss some possible continuations of our work. The methods employed in the present paper 
can be extended to implement the Wilsonian approach to gauge theories and quantum gravity, where some interesting attempts have already been made\cite{Gomis:1995jp,Branchina:2003ek,deAlwis:2017ysy}.
The same holds true for implementing conformal and/or scale invariance at the quantum level, where the present state of art is too poor. We plan to come to these issues in further studies. 

At the same time, extending the methods of the present work, we plan to pursue our investigation on the physical  mechanism that provides the tuning of the Higgs boson mass towards its experimental value.     

\vskip 20pt

\noindent
{\Large \bf Appendices}

\appendix

\renewcommand{\theequation}{A.\arabic{equation}}

\setcounter{equation}{0}

\section{Special functions}
The special functions $B(\alpha,\beta)$ and $\Gamma(z)$ of complex arguments $\alpha$, $\beta$ and $z$ are defined by
\begin{align}\label{betaf}
B\left( \alpha,\beta \right)&= \int^1_{0}dx\,x^{\alpha-1}(1-x)^{\beta-1}\qquad \mathrm{Re}\,\alpha\,,\mathrm{Re}\,\beta>0\\
\label{gammaf}
\Gamma\left(z\right)&= \int^\infty_{0}d\tau\,
\tau^{z-1}e^{-\tau} \qquad \qquad \,\,\,\,\, \mathrm{Re}\,z>0
\end{align}
where the conditions $\mathrm{Re}\,\alpha>0$, $\mathrm{Re}\,\beta>0$ and $\mathrm{Re}\,z>0$  guarantee the convergence of the integrals in (\ref{betaf}) and \eqref{gammaf}. The functions $B$ and $\Gamma$ satisfy the relations
\begin{align}\label{cgamma}
\Gamma(z+1)&=z \,\Gamma(z)\\ 
\label{cbeta}
B\left(\alpha,\beta\right)&=\frac{\Gamma\left(\alpha\right)\Gamma\left(\beta\right)}{\Gamma(\alpha+\beta)}\,,
\end{align}
and the inverse of $\Gamma(z)$ can be given with the help of the Weierstrass representation ($\gamma_{_E}$ is the Euler-Mascheroni constant):

\begin{equation}\label{Weier}
\Gamma(z)^{-1}=z\, e^{\gamma_{_E}\, z} \prod_{n=1}^{\infty} \left( 1+\frac{z}{n} \right)e^{-\frac{z}{n}} \qquad \qquad \mathrm{Re}\,z>0\,.
\end{equation}

The right hand side of \eqref{Weier} has zeros for $z=0,-1,-2,\dots$ 
(but these values are excluded by the condition $\mathrm{Re}\,z >0$), and is convergent for any $z$. 
The analytic extension 
$\overline \Gamma(z)$ 
of $\Gamma(z)$ is given by the inverse of the right hand side of \eqref{Weier}, and then it is defined for generic values of $z$, with the exception of the zeros of \eqref{Weier}. The property \eqref{cgamma} holds also for $\overline \Gamma(z)$.

The analytic extension $\overline B(\alpha,\beta)$  of $B(\alpha,\beta)$ is obtained with the help of \eqref{cbeta}, 
once the replacement  $\Gamma \to \overline\Gamma$ is made. Due to the properties of the function $\overline{\Gamma}(z)$ defined above, the function $\overline B(\alpha,\beta)$ is defined for generic complex values of 
$\alpha$ and $\beta$, excluding $\alpha,\beta= 0,-1,-2,\dots$

Another special function used in the text is the incomplete beta function $B_i(\alpha,\beta;x)$ defined as
\be \label{betafi}
\,\,\,\, B_i(\alpha,\beta; x)\equiv \int^x_{0}dy\,
y^{\alpha-1}(1-y)^{\beta-1}\qquad \qquad  {\rm Re}\,\alpha,\, {\rm Re}\,\beta >0,\, x\in \mathbb{R} 
\ee

It is not difficult to show that, 
when $|x|<1$, ${\rm Re}\,\alpha >0$, and ${\rm Re}\,\beta>0$,  the function $B_i(\alpha,\beta;x)$ satisfies the relation 
\begin{equation}\label{Bipropa}
B_i(\alpha,\beta;x)=\frac{x^\alpha}{\alpha}\,{}_2F_1\left( \alpha,1-\beta; \alpha+1; x \right),
\end{equation}
where  ${}_2F_1\left( a,b; c; x \right)$ is the hypergeometric function
\begin{equation}\label{hypefunct}
{}_2F_1\left( a,b; c; x \right) = \sum_{n=0}^\infty \frac{(a)_n (b)_n}{(c)_n} \frac{x^n}{n!}\,,
\end{equation}
and $(y)_n$ are the Pochammer symbols
\begin{equation}\label{pochy}
(y)_n\equiv\frac{\Gamma(y+n)}{\Gamma(y)}=y(y+1) \cdots (y+n-1)\,.
\end{equation}

From (\ref{hypefunct}) and (\ref{pochy}) we see that the right hand side 
of (\ref{Bipropa}) is defined for
any value of $\beta$ and for generic values of $\alpha$, 
(not only for ${\rm Re}\,\alpha >0$, and ${\rm Re}\,\beta>0$ as in \eqref{betafi}), 
with the exception of the values
$\alpha= 0, -1, -2, \dots$ 
With the help of (\ref{Bipropa}), the analytic extension  
$\overline B_i(\alpha,\beta;x)$ 
of $B_i(\alpha,\beta;x)$
is then defined in this larger domain.

\section{Propagator in $d$-dimensions. Proper-time}

\renewcommand{\theequation}{B.\arabic{equation}}

\setcounter{equation}{0}

In this appendix we show that results similar to those obtained in Section 3,
where a hard momentum cut-off is used, can be obtained when we consider a proper-time regularization for the loop integrals. To this end, rather than resorting to the example of the full effective potential as in Section 3, we consider only the one-loop two-point vertex function $\Gamma^{(2)}(0)$ in $d$-dimensions for zero external momenta
\begin{eqnarray}\label{Function2}
\Gamma^{(2)}(0)=m^2+ \delta m^2 +\frac{\lambda}{2}\mu^{4-d}\int \frac{d^dk}{(2\pi)^d}\frac{1}{k^2+m^2}\,.
\end{eqnarray}

Within the proper-time regularization, the loop integral in \eqref{Function2} is
\begin{eqnarray}\label{properInt}
\frac{1}{m^2}\int \frac{d^dk}{(2\pi)^d}\int^{\infty}_{m^2/\Lambda^2} d\tau\, 
e^{-\tau (k^2/m^2+1)}=\frac{(m^2)^{d/2-1}}{(4\pi)^{d/2}}\int^{\infty}_{m^2/\Lambda^2} d\tau\, 
\tau^{-d/2}\,e^{-\tau}\,.
\end{eqnarray}

Although $d$ in (\ref{properInt}) is a positive integer, the integral in the right hand side of this equation is convergent {\it for any complex value of} $d$. From the  definition of the lower incomplete Gamma function $\Gamma_i(z;u)$ (with $z\in \mathbb{C}$ and $u \in \mathbb{R}$)
\begin{equation}\label{lig}
\Gamma_i(z;u) \equiv  \int_u^\infty d \tau \, \tau^{z-1} \, e^{-\tau} \,,
\end{equation}
we have
\begin{equation}\label{incgamma}
\int^{\infty}_{m^2/\Lambda^2} d\tau\,
\tau^{-d/2} e^{-\tau}= \,\Gamma_i\left(1-\frac{d}{2};\frac{m^2}{\Lambda^2}\right)\,.
\end{equation}
Moreover, for ${\rm Re}\,d <2$, we have 
\begin{equation}
\Gamma_i\left(1-\frac{d}{2};\frac{m^2}{\Lambda^2}\right)=\Gamma\left(1-\frac{d}{2}\right)-\gamma\left(1-\frac{d}{2};\frac{m^2}{\Lambda^2}\right),
\end{equation}
where $\gamma(z,u)$ is the upper incomplete Gamma function (with ${\rm Re}\,z>0$ and $u \in \mathbb{R}$)
\begin{equation}\label{uig}
\gamma(z;u) \equiv  \int^u_0 d \tau \, \tau^{z-1} \, e^{-\tau} \,.
\end{equation}

This latter function satisfies the relation
\begin{equation}\label{gaincproperty}
\gamma(z,u)=\frac{u^z}{z} \, {}_1F_1(z,z+1;-u) \,,
\end{equation}
where ${}_1F_1(a,b;u)$ is the hypergeometric function
\begin{equation}\label{hypfunct1}
{}_1F_1\left( a,b; u \right) = \sum_{n=0}^\infty \frac{(a)_n}{(b)_n} \, \frac{u^n}{n!}\,,
\end{equation}
and $(x)_n$ are the Pochammer symbols defined in (\ref{pochy}).

The hypergeometric function ${}_1F_1(a,b;u)$ is defined for any complex value of $u$ (infinite convergence radius), for any complex value of $a$, and for generic complex values of $b$, excluding $b = 0,-1,-2,\dots$. The analytic extension $\overline \gamma(z;u)$ of $\gamma(z;u)$ is defined through the right hand side of \eqref{gaincproperty} 
\begin{equation}\label{gaincext}
\overline \gamma(z,u)=\frac{u^z}{z} \, {}_1F_1(z,z+1;-u) \qquad {\rm for}\, z \neq 0,-1,-2,\dots\,,
\end{equation}
and it is not difficult to see that, for any integer positive value of $d$,
\be\label{splittinggamma}
\int^{\infty}_{m^2/\Lambda^2} d\tau\,
\tau^{-d/2} e^{-\tau}=\Gamma_i\left(1-\frac{d}{2};\frac{m^2}{\Lambda^2}\right)=\lim_{z \to d}\left[\overline \Gamma\left(1-\frac{z}{2}\right)-
\overline \gamma\left(1-\frac{z}{2};\frac{m^2}{\Lambda^2}\right)\right] .
\ee

The reader might find convenient at this point to compare the above equation with the analogous Eq.\,\eqref{newsplitting0} of Section 3.
We already know that $\overline \Gamma \left(1-\frac z2\right)$ in \eqref{splittinggamma} has simple poles in $z=2,4,6,\dots$.  Moreover, expanding $\overline \gamma$ in powers of $\frac{m^2}{\Lambda^2}\ll1$ with the help of (\ref{hypfunct1}) and (\ref{gaincext}), we have
\begin{eqnarray}\label{gammaexp}
\overline \gamma\left(1-\frac{z}{2},\frac{m^2}{\Lambda^2}\right)=\frac{2}{2-z}
\left(\frac{m^2}{\Lambda^2}\right)^{\frac{2-z}{2}}
-\frac{2}{4-z}\left(\frac{m^2}{\Lambda^2}\right)^{\frac{4-z}{2}}
+\frac{1}{6-z}\left(\frac{m^2}{\Lambda^2}\right)^{\frac{6-z}{2}}
+\dots
\end{eqnarray} 
that shows that $\overline \gamma\left(1-\frac{z}{2},\frac{m^2}{\Lambda^2}\right)$ has simple poles in $z=2,4,6,\dots$ as $\overline \Gamma\left(1-\frac{z}{2}\right)$. 

By considering for the two-point function $\Gamma^{(2)}(0)$ in \eqref{Function2} a proper-time regularization, we have found a result in all similar to the one obtained in Section 3 for the effective potential, where we used a hard momentum cut-off.

Following the same approach of Section 3, from the proper-time regularization (that is a way of implementing the Wilsonian strategy with a smooth cut-off) we can again derive the DR rules.

Specifically, writing $z=4-\epsilon$ and expanding around $\epsilon=0$, from \eqref{properInt} and \eqref{splittinggamma} we have
\begin{eqnarray}\label{loopintegr2}
\left[\frac{\lambda}{2} \mu^{4-d}\,\frac{(m^2)^{d/2-1}}{(4\pi)^{d/2}}\int^{\infty}_{m^2/\Lambda^2} d\tau\, 
\tau^{-d/2}\,e^{-\tau} \right]_{d=4} =\lim_{\epsilon \to 0}\Big[C_1(4-\epsilon)-C_2(4-\epsilon) \Big]
\end{eqnarray}
where
\begin{align}\label{epssi3}
C_1(\epsilon)
&=\frac{\lambda m^2}{32\pi^2}\left(-\frac{2}{\epsilon}+\gamma_{_{E}}-\ln(4\pi)\right)+\frac{\lambda m^2}{32\pi^2}\left(\ln \frac{m^2}{\mu^2}-1\right)+\mathcal{O}(\epsilon)\,, \\
\label{ginexp}
C_2(\epsilon)
&=\frac{\lambda m^2}{32\pi^2}\left(-\frac{2}{\epsilon}-\ln(4\pi)\right)-\frac{\lambda m^2}{32\pi^2} \ln \frac{m^2}{\mu^2}\nonumber\\
&-\frac{\lambda \Lambda^2}{32\pi^2}
+\frac{\lambda m^2}{32\pi^2}\ln\frac{\Lambda^2}{m^2} +\mathcal{O}(\epsilon)
+\mathcal{O}(m^2/\Lambda^2)\,.
\end{align}

The similarity of Eqs.\,\eqref{epssi3} and \eqref{ginexp} with Eqs.\,\eqref{epsA1} and \eqref{epsA2} of Section 3 is evident, and the way to obtain the DR rules following the same path illustrated in Fig.\,\ref{Stepfigure} is immediately clear. The fine-tuning for the mass parameter, as explained in detail in Section 3, is hidden in the step 
$
\begin{tikzpicture}[baseline=-3]
\begin{feynman}
\vertex[draw, circle, minimum size=0.5cm,very thick] (m) at (0, 0) {\footnotesize{$1$}};
\diagram* {
	(m)
};
\end{feynman}
\end{tikzpicture}
\to
\begin{tikzpicture}[baseline=-3]
\begin{feynman}
\vertex[draw, circle, minimum size=0.5cm,very thick] (m) at (0, 0) {\footnotesize{$2$}};
\diagram* {
	(m)
};
\end{feynman}
\end{tikzpicture}
$\,, 
and again we see that DR is a way of implementing the Wilsonian calculation, incorporating the fine-tuning of the mass parameter. Naturally, again with reference to Fig.\,\ref{Stepfigure}, if we follow the path $
\begin{tikzpicture}[baseline=-3]
\begin{feynman}
\vertex[draw, circle, minimum size=0.5cm,very thick] (m) at (0, 0) {\footnotesize{$1$}};
\diagram* {
	(m)
};
\end{feynman}
\end{tikzpicture}
\to
\begin{tikzpicture}[baseline=-3]
\begin{feynman}
\vertex[draw, circle, minimum size=0.5cm,very thick] (m) at (0, 0) {\footnotesize{$3$}};
\diagram* {
	(m)
};
\end{feynman}
\end{tikzpicture}
$\,, 
we obtain 
\begin{align}\label{cc}
\left[\frac{\lambda}{2}\mu^{4-d}\frac{(m^2)^{d/2-1}}{(4\pi)^{d/2}}\int^{\infty}_{m^2/\Lambda^2} d\tau\, 
\tau^{-d/2}\,e^{-\tau}\right]_{d=4}=\frac{\lambda \Lambda^2}{32\pi^2}-\frac{\lambda m^2}{32\pi^2}\ln \frac{\Lambda^2}{m^2}+\frac{\lambda m^2}{32\pi^2}\left(\gamma_{_{E}}-1\right)
\end{align}
and the fine-tuning has to be implemented in the usual manner.


\begin{thebibliography}{100}

\bibitem{ATLAS:2012yve}
G.~Aad \textit{et al.} [ATLAS],
``Observation of a new particle in the search for the Standard Model Higgs boson with the ATLAS detector at the LHC",
Phys. Lett. B \textbf{716}, 1-29 (2012).
\bibitem{CMS:2012qbp}
S.~Chatrchyan \textit{et al.} [CMS],
``Observation of a New Boson at a Mass of 125 GeV with the CMS Experiment at the LHC",
Phys. Lett. B \textbf{716}, 30-61 (2012).


\bibitem{Giudice:2013yca}
G.~F.~Giudice,
``Naturalness after LHC8",
PoS \textbf{EPS-HEP2013}, 163 (2013).


\bibitem{Susskind:1978ms}
L.~Susskind,
``Dynamics of Spontaneous Symmetry Breaking in the Weinberg-Salam Theory",
Phys. Rev. D \textbf{20}, 2619-2625 (1979).

\bibitem{deGouvea:2014xba}
A.~de Gouvea, D.~Hernandez and T.~M.~P.~Tait,
``Criteria for Natural Hierarchies",
Phys. Rev. D \textbf{89}, no.11, 115005 (2014).

\bibitem{Glazek:1994qc}
S.~D.~Glazek and K.~G.~Wilson,
``Perturbative renormalization group for Hamiltonians",
Phys. Rev. D \textbf{49}, 4214-4218 (1994).

\bibitem{tHooft:2011aa}
G.~'t Hooft,
``A class of elementary particle models without any adjustable real parameters",
Found. Phys. \textbf{41}, 1829-1856 (2011).

\bibitem{tHooft:2016uxd}
G.~'t Hooft,
``Local conformal symmetry in black holes, standard model, and quantum gravity",
Int. J. Mod. Phys. D \textbf{26}, no.03, 1730006 (2016).

\bibitem{Meissner:2006zh}
K.~A.~Meissner and H.~Nicolai,
``Conformal Symmetry and the Standard Model",
Phys. Lett. B \textbf{648}, 312-317 (2007).

\bibitem{Meissner:2007xv}
K.~A.~Meissner and H.~Nicolai,
``Effective action, conformal anomaly and the issue of quadratic divergences",
Phys. Lett. B \textbf{660}, 260-266 (2008).

\bibitem{Foot:2007iy}
R.~Foot, A.~Kobakhidze, K.~L.~McDonald and R.~R.~Volkas,
``A Solution to the hierarchy problem from an almost decoupled hidden sector within a classically scale invariant theory",
Phys. Rev. D \textbf{77}, 035006 (2008).

\bibitem{Bezrukov:2007ep}
F.~L.~Bezrukov and M.~Shaposhnikov,
``The Standard Model Higgs boson as the inflaton",
Phys. Lett. B \textbf{659}, 703-706 (2008).

\bibitem{Shaposhnikov:2008xi}
M.~Shaposhnikov and D.~Zenhausern,
``Quantum scale invariance, cosmological constant and hierarchy problem",
Phys. Lett. B \textbf{671}, 162-166 (2009).

\bibitem{Alexander-Nunneley:2010tyr}
L.~Alexander-Nunneley and A.~Pilaftsis,
``The Minimal Scale Invariant Extension of the Standard Model,
JHEP \textbf{09}, 021 (2010).

\bibitem{Boyle:2011fq}
L.~Boyle, S.~Farnsworth, J.~Fitzgerald and M.~Schade,
``The Minimal Dimensionless Standard Model (MDSM) and its Cosmology",
arXiv:1111.0273 [hep-ph].

\bibitem{Carone:2013wla}
C.~D.~Carone and R.~Ramos,
``Classical scale-invariance, the electroweak scale and vector dark matter",
Phys. Rev. D \textbf{88}, 055020 (2013).

\bibitem{Farzinnia:2013pga}
A.~Farzinnia, H.~J.~He and J.~Ren,
``Natural Electroweak Symmetry Breaking from Scale Invariant Higgs Mechanism",
Phys. Lett. B \textbf{727}, 141-150 (2013).

\bibitem{Kawamura:2013kua}
Y.~Kawamura,
``Naturalness, Conformal Symmetry and Duality",
PTEP \textbf{2013}, no.11, 113B04 (2013).

\bibitem{Bars:2013yba}
I.~Bars, P.~Steinhardt and N.~Turok,
``Local Conformal Symmetry in Physics and Cosmology",
Phys. Rev. D \textbf{89}, no.4, 043515 (2014).

\bibitem{Heikinheimo:2013fta}
M.~Heikinheimo, A.~Racioppi, M.~Raidal, C.~Spethmann and K.~Tuominen,
``Physical Naturalness and Dynamical Breaking of Classical Scale Invariance",
Mod. Phys. Lett. A \textbf{29}, 1450077 (2014).

\bibitem{Salvio:2014soa}
A.~Salvio and A.~Strumia,
``Agravity",
JHEP \textbf{06}, 080 (2014).

\bibitem{Steele:2013fka}
T.~G.~Steele, Z.~W.~Wang, D.~Contreras and R.~B.~Mann,
``Viable dark matter via radiative symmetry breaking in a scalar singlet Higgs portal extension of the standard model",
Phys. Rev. Lett. \textbf{112}, no.17, 171602 (2014).

\bibitem{Bezrukov:2014ipa}
F.~Bezrukov, J.~Rubio and M.~Shaposhnikov,
``Living beyond the edge: Higgs inflation and vacuum metastability",
Phys. Rev. D \textbf{92}, no.8, 083512 (2015).

\bibitem{Guo:2014bha}
J.~Guo and Z.~Kang,
``Higgs Naturalness and Dark Matter Stability by Scale Invariance",
Nucl. Phys. B \textbf{898}, 415-430 (2015).

\bibitem{Ghilencea:2015mza}
D.~M.~Ghilencea,
``Manifestly scale-invariant regularization and quantum effective operators",
Phys. Rev. D \textbf{93}, no.10, 105006 (2016).

\bibitem{Ghilencea:2016ckm}
D.~M.~Ghilencea, Z.~Lalak and P.~Olszewski,
``Two-loop scale-invariant scalar potential and quantum effective operators",
Eur. Phys. J. C \textbf{76}, no.12, 656 (2016).

\bibitem{Wang:2015cda}
Z.~W.~Wang, T.~G.~Steele, T.~Hanif and R.~B.~Mann,
``Conformal Complex Singlet Extension of the Standard Model: Scenario for Dark Matter and a Second Higgs Boson",
JHEP \textbf{08}, 065 (2016).

\bibitem{Ghilencea:2016dsl}
D.~M.~Ghilencea, Z.~Lalak and P.~Olszewski,
``Standard Model with spontaneously broken quantum scale invariance,
Phys. Rev. D \textbf{96}, no.5, 055034 (2017).

\bibitem{Oda:2018zth}
I.~Oda,
``Planck and Electroweak Scales Emerging from Conformal Gravity",
Eur. Phys. J. C \textbf{78}, no.10, 798 (2018).

\bibitem{Mooij:2018hew}
S.~Mooij, M.~Shaposhnikov and T.~Voumard,
``Hidden and explicit quantum scale invariance",
Phys. Rev. D \textbf{99}, no.8, 085013 (2019).

\bibitem{Farina:2013mla}
M.~Farina, D.~Pappadopulo and A.~Strumia,
``A modified naturalness principle and its experimental tests",
JHEP \textbf{08}, 022 (2013).

\bibitem{Brivio:2017vri}
I.~Brivio and M.~Trott,
``The Standard Model as an Effective Field Theory",
Phys. Rept. \textbf{793}, 1-98 (2019).

\bibitem{DeAlwis:2021gou}
S.~P.~De Alwis,
``Wilsonian Effective Field Theory and String Theory",
arXiv:2103.13347 [hep-th].

\bibitem{deAlwis:2021zab}
S.~P.~de Alwis,
``Radiative Generation of dS from AdS",
arXiv:2110.06967 [hep-th].

\bibitem{Alexandre:1997gj}
J.~Alexandre, V.~Branchina and J.~Polonyi,
``Global renormalization group",
Phys. Rev. D \textbf{58}, 016002 (1998).

\bibitem{Alexandre:1998ts}
J.~Alexandre, V.~Branchina and J.~Polonyi,
``Instability induced renormalization",
Phys. Lett. B \textbf{445}, 351-356 (1999).

\bibitem{Coleman:1973jx}
S.~R.~Coleman and E.~J.~Weinberg,
``Radiative Corrections as the Origin of Spontaneous Symmetry Breaking",
Phys. Rev. D \textbf{7}, 1888-1910 (1973).

\bibitem{Clark:1992jr}
T.~E.~Clark, B.~Haeri and S.~T.~Love,
``Wilson renormalization group analysis of theories with scalars and fermions",
Nucl. Phys. B \textbf{402}, 628-656 (1993).

\bibitem{Krajewski:2014vea}
T.~Krajewski and Z.~Lalak,
``Fine-tuning and vacuum stability in the Wilsonian effective action",
Phys. Rev. D \textbf{92}, no.7, 075009 (2015).

\bibitem{Shaposhnikov:2009pv}
M.~Shaposhnikov and C.~Wetterich,
``Asymptotic safety of gravity and the Higgs boson mass'',
Phys. Lett. B \textbf{683}, 196-200 (2010).

\bibitem{Veltman:1980mj}
M.~J.~G.~Veltman,
``The Infrared - Ultraviolet Connection",
Acta Phys. Polon. B \textbf{12}, 437 (1981).

\bibitem{Holthausen:2013ota}
M.~Holthausen, J.~Kubo, K.~S.~Lim and M.~Lindner,
``Electroweak and Conformal Symmetry Breaking by a Strongly Coupled Hidden Sector",
JHEP \textbf{12} (2013), 076.

\bibitem{Hamada:2012bp}
Y.~Hamada, H.~Kawai and K.~y.~Oda,
``Bare Higgs mass at Planck scale",
Phys. Rev. D \textbf{87} 5, 053009 (2013)
[erratum: Phys. Rev. D \textbf{89} 5].

\bibitem{Jones:2013aua}
D.~R.~T.~Jones,
``Comment on \textquotedblleft{}Bare Higgs mass at Planck scale\textquotedblright{}",
Phys. Rev. D \textbf{88} 9, 098301 (2013).

\bibitem{Chankowski:2014fva}
P.~H.~Chankowski, A.~Lewandowski, K.~A.~Meissner and H.~Nicolai,
``Softly broken conformal symmetry and the stability of the electroweak scale'',
Mod. Phys. Lett. A \textbf{30}, no.02, 1550006 (2015).

\bibitem{Ford:1992mv}
C.~Ford, D.~R.~T.~Jones, P.~W.~Stephenson and M.~B.~Einhorn,
``The Effective potential and the renormalization group",
Nucl. Phys. B \textbf{395}, 17-34 (1993).

\bibitem{Holthausen:2011aa}
M.~Holthausen, K.~S.~Lim and M.~Lindner,
``Planck scale Boundary Conditions and the Higgs Mass'',
JHEP \textbf{02}, 037 (2012).

\bibitem{Iso:2012jn}
S.~Iso and Y.~Orikasa,
``TeV Scale B-L model with a flat Higgs potential at the Planck scale: In view of the hierarchy problem",
PTEP \textbf{2013} (2013), 023B08.

\bibitem{Ibanez:2013gf}
L.~E.~Ibanez and I.~Valenzuela,
``The Higgs Mass as a Signature of Heavy SUSY",
JHEP \textbf{05} (2013), 064.

\bibitem{Masina:2013wja}
I.~Masina and M.~Quiros,
``On the Veltman Condition, the Hierarchy Problem and High-Scale Supersymmetry",
Phys. Rev. D \textbf{88} (2013), 093003.

\bibitem{Hashimoto:2013hta}
M.~Hashimoto, S.~Iso and Y.~Orikasa,
``Radiative symmetry breaking at the Fermi scale and flat potential at the Planck scale",
Phys. Rev. D \textbf{89} (2014) no.1, 016019.

\bibitem{Haba:2013lga}
N.~Haba, K.~Kaneta and R.~Takahashi,
``Planck scale boundary conditions in the standard model with singlet scalar dark matter",
JHEP \textbf{04} (2014), 029.

\bibitem{Hashimoto:2014ela}
M.~Hashimoto, S.~Iso and Y.~Orikasa,
``Radiative symmetry breaking from flat potential in various U(1)' models",
Phys. Rev. D \textbf{89} (2014) no.5, 056010.

\bibitem{Sirlin:1985ux}
A.~Sirlin and R.~Zucchini,
``Dependence of the Quartic Coupling $\overline h_{\overline{MS}}(M)$ on $m_H$ and the Possible Onset of New Physics in the Higgs Sector of the Standard Model'',
Nucl. Phys. B \textbf{266}, 389-409 (1986).

\bibitem{Wetterich:1981ir}
C.~Wetterich,
``Gauge hierarchy due to strong interactions?",
Phys. Lett. B \textbf{104}, 269-276 (1981).

\bibitem{Wetterich:1983bi}
C.~Wetterich,
``Fine Tuning Problem and the Renormalization Group",
Phys. Lett. B \textbf{140}, 215-222 (1984).

\bibitem{Wetterich:1990an}
C.~Wetterich,
``Quadratic Renormalization of the Average Potential and the Naturalness of Quadratic Mass Relations for the Top Quark",
Z. Phys. C \textbf{48}, 693-705 (1990).

\bibitem{Wetterich:1991be}
C.~Wetterich,
``The Average action for scalar fields near phase transitions",
Z. Phys. C \textbf{57}, 451-470 (1993).

\bibitem{Bornholdt:1992up}
S.~Bornholdt and C.~Wetterich,
``Selforganizing criticality, large anomalous mass dimension and the gauge hierarchy problem",
Phys. Lett. B \textbf{282}, 399-405 (1992).

\bibitem{Wetterich:2011aa}
C.~Wetterich,
``Where to look for solving the gauge hierarchy problem?",
Phys. Lett. B \textbf{718}, 573-576 (2012).

\bibitem{Wetterich:2016uxm}
C.~Wetterich and M.~Yamada,
``Gauge hierarchy problem in asymptotically safe gravity--the resurgence mechanism",
Phys. Lett. B \textbf{770}, 268-271 (2017)

\bibitem{Pawlowski:2018ixd}
J.~M.~Pawlowski, M.~Reichert, C.~Wetterich and M.~Yamada,
``Higgs scalar potential in asymptotically safe quantum gravity",
Phys. Rev. D \textbf{99}, no.8, 086010 (2019).

\bibitem{Mooij:2021ojy}
S.~Mooij and M.~Shaposhnikov,
``QFT without infinities and hierarchy problem",
arXiv:2110.05175 [hep-th].

\bibitem{Mooij:2021lbc}
S.~Mooij and M.~Shaposhnikov,
``Finite Callan-Symanzik renormalisation for multiple scalar fields", arXiv:2110.15925 [hep-th].

\bibitem{Lenshina:2020edt}
N.~D.~Lenshina, A.~A.~Radionov and F.~V.~Tkachov,
``MS$^{4}$: An Alternative to the Bogolyubov\textendash{}Parasiuk\textendash{}Hepp\textendash{}Zimmermann (BPHZ) Theory",
Phys. Part. Nucl. \textbf{51}, no.4, 567-571 (2020).

\bibitem{Hadamard1}
J. Hadamard, \vv Lectures on Cauchy's problem in linear partial differential equations", Dover Phoenix editions, Dover Publications, New York (1923).

\bibitem{Hadamard2}
J. Hadamard, \vv Le problème de Cauchy et les équations aux dérivées partielles linéaires hyperboliques", Paris: Hermann \& Cie (1932).

\bibitem{Callan:1970yg}
C.~G.~Callan, Jr.,
``Broken scale invariance in scalar field theory",
Phys. Rev. D \textbf{2}, 1541-1547 (1970).

\bibitem{Symanzik:1970rt}
K.~Symanzik,
``Small distance behavior in field theory and power counting",
Commun. Math. Phys. \textbf{18}, 227-246 (1970).

\bibitem{Blaer:1974foy}
A.~S.~Blaer and K.~Young,
``Field theory renormalization using the Callan-Symanzik equation",
Nucl. Phys. B \textbf{83}, 493-514 (1974).

\bibitem{Callan:1975vs}
C.~G.~Callan, Jr.,
``Introduction to Renormalization Theory",
Conf. Proc. C \textbf{7507281}, 41-77 (1975).

\bibitem{Gomis:1995jp}
J.~Gomis and S.~Weinberg,
``Are nonrenormalizable gauge theories renormalizable?",
Nucl. Phys. B \textbf{469}, 473-487 (1996).

\bibitem{Branchina:2003ek}
V.~Branchina, K.~A.~Meissner and G.~Veneziano,
``The Price of an exact, gauge invariant RG flow equation,''
Phys. Lett. B \textbf{574}, 319-324 (2003).

\bibitem{deAlwis:2017ysy}
S.~P.~de Alwis,
``Exact RG Flow Equations and Quantum Gravity,''
JHEP \textbf{03}, 118 (2018). 

\end{thebibliography}
\end{document}